\documentclass[aps,showpacs,amsmath,amssymb]{revtex4}

\usepackage{latexsym}
\usepackage{graphicx}
\usepackage{dcolumn}
\usepackage{bm}
\usepackage{hyperref}

\newcommand{\beq}{\begin{equation}}
\newcommand{\eeq}{\end{equation}}
\newcommand{\beqa}{\begin{eqnarray}}
\newcommand{\eeqa}{\end{eqnarray}}
\newcommand{\beqar}{\begin{eqnarray*}}
\newcommand{\eeqar}{\end{eqnarray*}}
\newcommand{\bra}[1]{\mbox{$\left\langle{#1}\right|$}}
\newcommand{\ket}[1]{\mbox{$\left|{#1}\right\rangle$}}

\def\I{{\rm i}}
\def\d{{\rm d}}
\def\e{{\rm e}}
\def\Tr{{\rm Tr}}

\newcounter{saveeqn}

\begin{document}

\title{Quantum mechanics in general quantum systems (III): \\ open system dynamics}
\author{An Min Wang}\email{anmwang@ustc.edu.cn}
\affiliation{Quantum Theory Group, Department of Modern Physics,
University of Science and Technology of China, Hefei, 230026,
P.R.China}

\begin{abstract}

We investigate the exact solution, perturbation theory and master
equation of open system dynamics based on our serial studies on
quantum mechanics in general quantum systems [An Min Wang,
quant-ph/0611216, quant-ph/0611217]. In a system-environment
separated representation, a general and explicit solution of open
system dynamics is obtained, and it is an exact solution since it
includes all order approximations of perturbation. In terms of the
cut-off approximation of perturbation and our improved scheme of
perturbation theory, the improved form of the perturbed solution of
open systems absorbing the partial contributions from the high order
even all order approximations is deduced. Moreover, only under the
factorizing initial condition, the exact master equation including
all order approximations is proposed. Correspondingly, the perturbed
master equation and its improved form different from the existed
master equation are given. In special, the Redfield master equation
is derived out without using Born-Markov approximation. The solution
of open system dynamics in the Milburn model is also gained. As
examples, Zurek model of two-state open system and its extension
with two transverse fields are studied.

\end{abstract}


\maketitle

\section{Introduction}\label{sec1}

A realistic quantum system is never isolated, but is immersed in the
surrounding environment (alias bath, reservoir) and interacts
continuously with it. Such a system without ignorable coupling to
the environment can be called open quantum system. Generally, the
environment consists of a huge number of degrees of freedom, it is
even whole outside world (universe) of the concerning open quantum
system. In fact, we might not know the exact state of the outside
world, having only some statistical information to describe it.
However, we are really interested in a reliable and effective theory
of open system dynamics under the influence of its environment.

The basic idea of quantum theory of open systems is thought of as an
interesting open system and its surrounding environment form a total
composite system, or vis versa, a composite system can be decomposed
into an interesting open system and a surrounding environment. The
key matters of quantum theory of open systems are to determine the
interaction between the open system and its environment and build
the physical models of the open system and its environment. Open
system dynamics is just a law of this open quantum system how to
evolute with time and its solution at any given time.

Open quantum system and its dynamics are very important for many
interesting quantum theory branches such as quantum optics
\cite{Carmichael,Plenio}, condensed matter theory, quantum
information and computing \cite{Nielson,Perskill}, more concretely,
quantum decoherence, quantum measurement \cite{Zurek,Schlosshauer},
quantum dissipation \cite{Weiss,Strunz}, quantum transport
\cite{Haug}, quantum chaos \cite{Haake} et. al. Study of open system
dynamics is helpful for understanding some very essential problems
in physics, for example, the transition from quantum to classical
world.

A variety of different formal techniques have been developed and
used in dealing with open quantum systems. From the above reviews
and books, the interested readers may get them. Here, we intend to
start with ``the first principle" of quantum mechanics, that is, the
Schr\"odinger equation or the von Neumann equation, and then try to
build a theoretical formulism including the general and explicit
forms of motion equation, dynamical solution, and perturbation
theory of open systems.

It is clear that such a ``first principle" scheme might be not
suitable to the cases when one cannot clearly know the environment
model and/or the system-environment coupling form since the
environment is too huge and too complicated. However, our
conclusions might be helpful for building the models of such some
systems. Moreover, one of possible ways to avoid this difficulty is
to use the Milburn dynamics \cite{Milburn}. That is, the environment
is separated into near- and remote two parts, the Hamiltonians of
the near environment (often with finite degree of freedoms) and the
coupling to the interesting open system are assumed to be clearly
known, but the influence of the remote environment on the
interesting system is incarnated by an extra term in Milburn motion
equation compared with the von Neumann motion equation. Similarly,
we successfully obtain the general and explicit solution of Milburn
dynamics of the interesting system according our scheme.

Because of the dissipative nature of open systems, we must turn to
the density matrix for a proper description whatever the initial
state is pure or mixed. Actually, we are interested in the
properties of open systems only, it will be appropriate to study the
reduced density matrix evolution with time or motion equation or its
solution. Here, the reduced density matrix describing the open
systems is obtained by tracing out the degrees of freedom of the
environment from the total (system plus environment) density matrix.

Due to the system and environment being entangled generally in
system evolution with time, directly solving Schr\"odinger equation
or von Neumann equation of the total system is a formidable task by
using the existed methods. Traditionally, this problem is studied by
perturbation theory in system-environment coupling scheme. Ones
often take the interaction between the system and its environment as
a perturbed part and then use the interaction picture to derive out
the master equation of open systems via some physical approximations
such as Born-Markov ones and the others. If an open system is
exactly solvable, the coupling $H_{\rm SE}$ is weak, the evolution
time is short enough, and the used physical approximations are
indeed appropriate, this has been proved to be an effective method.
However, when the above conditions do not satisfied sufficiently,
the problem gets complicated and perhaps leads to some difficulties,
although some formal techniques have been developed and used in
order to overcome some possible shortcomings. For generality and
reliability in theory, we feel that we have to consider whether
these approximations are necessary, if without these approximations,
can we obtain the formulism of open system dynamics? The conclusions
obtained here answer these problems.

In this paper, we will provide the amelioration of the existed
scheme of open system dynamics and try to build a theoretical
formulism using our recent investigations on quantum mechanics in
general quantum systems \cite{My1,My2}. We first obtain the exact
solution of open systems including all order approximations of
perturbation and then give the improved form of perturbed solution
of open systems absorbing the partial contributions from the high
order even all order approximations of perturbation. Only under the
factorizing initial condition, we derive out the exact master
equation and its perturbed form via the standard cut-off
approximation of perturbation. Moreover, we propose the improved
form of perturbed master equation. In special, based on our master
equation, we re-deduce the Redfield master equation without using
Born-Markov approximation, and we point out the differences between
our master equation and existed ones. We also obtain the solution of
open system dynamics in the Milburn model. In order to illustrate
our open system dynamics, we study Zurek model of two-state open
system and its extension with two transverse fields. We are sure
that our open system dynamics can be used to more open systems since
its generality and clearness, and its calculations are simpler and
more efficient, its results are more accurate and more reliable than
the existed scheme.

This paper is arranged as follows: besides Sec. \ref{sec1} is an
introduction, in Sec. \ref{sec2}, by virtue of a system-environment
separated representation, we first obtain a general and explicit
solution of open systems including all order approximations; in Sec.
\ref{sec3}, we gain the improved form of perturbed solution of open
systems, which absorbs the partial contributions from the high order
even all order approximations of perturbation; in Sec. \ref{sec4} we
deduce the exact master equation of open systems only under the
factorizing initial condition; in Sec. \ref{sec5} we get the
perturbed form of our master equation and its amelioration; in Sec.
\ref{sec6}, based on our master equation, we re-deduce Redfield
master equation without using Born-Markov approximation, and we
point out the differences between our master equation and existed
ones; in Sec. \ref{sec7}, we obtain the solution of open system
dynamics in the Milburn model for the Milburn-type closed
total-systems. This implies that our solution and methods are
applicable to more general open systems; in Sec. \ref{sec8}, we
study Zurek model of two-state open system and its extension with
two transverse fields; In Sec. \ref{sec9}, we summarize our
conclusions and give some discussions.

\section{General and explicit solution of open system
dynamics}\label{sec2}

In this section, we will derive out a general and explicit solution
of open systems by using our recent work of exact solution in
general quantum systems \cite{My1}.

As is well-known, if assuming that the interesting open quantum
system and its environment are taken as a closed (or isolated)
larger composite system, that is, a total system, we can think that
this total system obeys the Schr\"odinger equation or the von
Neumann equation, respectively, for a pure state
$\ket{\Psi_{tot}(t)}$ or a mixed state ${\rho}_{tot}(t)$, that is
\beqa\label{firstse} -\I\frac{\partial}{\partial
t}\ket{\Psi_{tot}(t)}&=&H_{tot}\ket{\Psi_{tot}(t)},\\
\label{firstvne}
\dot{\rho}_{tot}(t)&=&-\I\left[H_{tot},\rho_{tot}(t)\right].\eeqa
where the total system Hamiltonian $H_{tot}$ that we consider here
is made of the sum of the interesting open system Hamiltonian
$H_{\rm S}$ and its surrounding environment Hamiltonian $H_{\rm E}$
plus an interaction $H_{\rm SE}$ between the system and the
environment, that is \beq
H_{tot}=H_{\rm{S}}+H_{\rm{E}}+H_{\rm{SE}}.\eeq Note that the total
system Hilbert space $\mathcal{H}_{tot}$ is defined by the direct
product $\mathcal{H}_{\rm S}\otimes \mathcal{H}_{\rm E}$ of open
system Hilbert space $\mathcal{H}_{\rm S}$ and its environment
Hilbert space $\mathcal{H}_{\rm E}$. Here and in the following, we
will discuss time-independent Hamiltonian and we have taken
$\hbar=1$ for simplicity.

In an open system dynamics, a key difficulty to lead to the problem
becomes intractable is that there is the interaction between the
open system and its environment with huge degree of freedom. With
time evolution, the open system inevitably entangles with its
environment. Therefore, we starts from a system-environment
separated representation (SESR). This representation is beneficial
for obtaining the general and explicit solution of open system
dynamics as well as proposing the improved scheme of perturbed
theory \cite{My2}, because in the SESR we can conveniently trace off
the degree of freedom of environment. Introducing the SESR is a
simple and natural idea, and we will see it is also very useful. To
this purpose, we first divide the $H_{tot}$ into two parts \beq
H_{tot}={H}_{tot0}+{H}_{tot1}, \eeq and, without loss of generality,
we denote \beqa\label{H0form} {H}_{tot0}&=&H_{\rm S0}+H_{\rm E0}+
H_{\rm SE0},\quad \label{H1form} {H}_{tot1}=H_{\rm S1}+H_{\rm E1}+
H_{\rm SE1}.\eeqa It is clear that \beqa H_{\rm S0}&=&h_{\rm
S0}\otimes I_{\rm E},\quad H_{\rm E0}=I_{\rm S}\otimes h_{\rm
E0},\eeqa while we need the coupling Hamiltonian with the following
form \beqa H_{\rm SE0}&=&\sum_{m,n}c_{mn} S_{m0}\otimes B_{n0}.
\eeqa It is general enough if we do not restrict the forms of
$S_{m0}$ and $B_{n0}$. In the above expressions, the total Hilbert
space is $\mathcal{H}_{tot}=\mathcal{H}_{\rm S}\otimes
\mathcal{H}_{\rm E}$, $I_{\rm S}$ and $I_{\rm E}$ are, respectively,
the identity operators in $\mathcal{H}_{\rm S}$ and $
\mathcal{H}_{\rm E}$, and $c_{mn}$ are coupling constants between
the open system and its environment. Note that $H_{\rm S0}$ and
$H_{\rm E0}$ are always hermitian as usual. In addition, we need
$H_{\rm{SE}0}$ be also necessarily hermitian. In fact, because
$H_{\rm S0}$ and $H_{\rm E0}$ commute, the SESR always exists. The
aim to add $H_{\rm SE0}$ is to obtain better precision and to
simplify the perturbed part when passing to perturbation theory. It
must be pointed out that the general principle to divide $H$ into
two parts is to let the terms as more as possible belong to ${H}_0$
but the precondition is that there exist the commuting relations:
\beq [h_{\rm S0},S_{m0}], \quad [h_{\rm E0},\sum_{n}
c_{mn}B_{n0}]=0, \quad \mbox{or}\quad [h_{\rm S0},\sum_m
c_{mn}S_{m0}], \quad [h_{\rm E0},B_{n0}]=0. \eeq Moreover, that the
eigenvalue problem of $H_{tot0}$ is solvable. In fact, this
solvability implies that $h_{\rm S0}$ and $h_{\rm E0}$ are solvable,
then $h_{\rm S0}$ and $S_m$, $h_{\rm E0}$ and $\sum_{n} c_{mn}B_n$
have the common eigenvectors, or $h_{\rm S0}$ and
$\sum_{m}c_{mn}S_m$, $h_{\rm E0}$ and $B_n$ have the common
eigenvectors i.e, we have, respectively, \beqa h_{\rm S0}
\ket{\phi^{\gamma}}&=&E_{\gamma}\ket{\phi^{\gamma}},\quad S_{m0}
\ket{\phi^{\gamma}}=s_{m\gamma}\ket{\phi^{\gamma}},\quad h_{\rm E0}
\ket{\chi_v}=\varepsilon_v\ket{\chi_v}, \quad \sum_{n} c_{mn}B_{n0}
\ket{\chi_v}=r_{mv}\ket{\chi_v} \eeqa and \beqa h_{\rm S0}
\ket{\phi^{\gamma}}&=&E_{\gamma}\ket{\phi^{\gamma}},\quad
\sum_{m}c_{mn}S_{m0}
\ket{\phi^{\gamma}}=s_{n\gamma}\ket{\phi^{\gamma}},\quad h_{\rm E0}
\ket{\chi_v}=\varepsilon_v\ket{\chi_v}, \quad B_{n0}
\ket{\chi_v}=r_{nv}\ket{\chi_v}. \eeqa

They indicate that the eigenvectors of $H_{tot0}$, or the common
eigenvectors of $H_{\rm S0}$, $H_{\rm E0}$ and $H_{\rm SE0}$ are
\beq \ket{\Phi^{\gamma
v}}=\ket{\phi^{\gamma}}\otimes\ket{\chi^{v}},\eeq which span a
separate representation of the system and the environment, and it is
clear that \beq H_{tot0}\ket{\Phi^{\gamma v}}=E_{\gamma
v}\ket{\Phi^{\gamma v}},\eeq \beq E_{\gamma v}=
E_{\gamma}+\varepsilon_v+\sum_{m} s_{m\gamma} r_{mv}\quad \mbox{or}
\quad E_{\gamma v}= E_{\gamma}+\varepsilon_v+\sum_{n} s_{n\gamma}
r_{nv}. \eeq

It must be emphasized that the principle of Hamiltonian split is not
just the best solvability in more general cases. If the cut-off
approximation of perturbation is necessary, it requires that the
off-diagonal elements of the perturbing Hamiltonian $H_{tot1}$
matrix in the SESR is small enough compared with the diagonal
elements of $H_{tot}=H_{tot0}+H_{tot1}$ matrix in the same
representation according to our improved scheme of perturbation
theory. In addition, if there are the degeneracies, the Hamiltonian
split is also restricted by the condition that the degeneracies can
be completely removed via the usual diagonalization procedure of the
degenerate subspaces and our Hamiltonian redivision, or specially,
if the remained degeneracies are allowed, it requires that the
off-diagonal elements of the perturbing Hamiltonian matrix between
any two degenerate levels are always vanishing, in order to let our
improved scheme of perturbation theory work well \cite{My2}. As an
example, it has been studied in Sec. \ref{sec8}.

From the formal solution of the von Neumann equation of the total
system \beq \label{gsrhos}\rho_{tot}(t)=\e^{-\I H_{tot}
t}\rho_{tot}(0)\e^{\I H_{tot} t},\eeq and our expression of the time
evolution operator \cite{My1} \beqa \e^{-\I H_{tot}
t}&=&\sum_{l=0}^\infty \mathcal{A}_l(t),\eeqa it immediately follows
that the solution of total system density matrix with time evolution
is \beq \rho_{tot}(t)=\sum_{k,l=0}^\infty
\mathcal{A}_k(t)\rho_{tot}(0)\mathcal{A}_l(-t)=\sum_{k,l=0}^\infty
\mathcal{A}_k(t)\rho_{tot}(0)\mathcal{A}^\dagger_l(t).\eeq

In the SESR, we have \beqa \label{gestos}
\rho_{tot}(t)&=&\sum_{\beta, u,\beta^\prime,u^\prime}\sum_{\gamma,
v,\gamma^\prime,v^\prime}\sum_{k,l=1}^\infty
A_k^{\gamma\beta}(t)\rho^{\beta u,\beta^\prime
u^\prime}(0)A_l^{\beta^\prime u^\prime, \gamma^\prime
v^\prime}(-t)\ket{\Phi^{\gamma v}}\bra{\Phi^{\gamma^\prime
v^\prime}}\\
&=& \sum_{\beta, u,\beta^\prime,u^\prime}\sum_{\gamma,
v,\gamma^\prime,v^\prime}\sum_{k,l=1}^\infty
A_k^{\gamma\beta}(t)\rho^{\beta u,\beta^\prime
u^\prime}(0)A_l^{\beta^\prime u^\prime, \gamma^\prime
v^\prime}(-t)\left[\ket{\phi^{\gamma}}\bra{\phi^{\gamma^\prime}}\right]
\otimes \left[\ket{\chi^v}\bra{\chi^{v^\prime}}\right],\eeqa where
\beqa A_l^{\gamma v,\gamma^\prime v^\prime}(t)&=&\bra{\Phi^{\gamma
v}}\mathcal{A}_l(t)\ket{\Phi^{\gamma^\prime v^\prime}},\\
\rho_{tot}^{\gamma v,\gamma^\prime v^\prime}(0)&=&\bra{\Phi^{\gamma
v}}\rho_{tot}(0)\ket{\Phi^{\gamma^\prime v^\prime}}.\eeqa In Ref.
\cite{My1}, we have found the explicit forms of $\mathcal{A}_l(t)$.
In the SESR, they read \beqa \label{Aldefinition} A_0^{\gamma
v,\gamma^\prime v^\prime}(t)&=&\e^{-\I E_{\gamma v}
t}\delta_{\gamma\gamma^\prime}\delta_{vv^\prime},\\ A_l^{\gamma
v,\gamma^\prime
v^\prime}(t)&=&\sum_{\gamma_1,\cdots,\gamma_{l+1}}\sum_{v_{\gamma_1},\cdots,v_{\gamma_{l+1}}}\left[
\sum_{i=1}^{l+1}(-1)^{i-1}\frac{\e^{-\I E_{\gamma_iv_i}
t}}{d_i(E[\gamma v,l])}\right]\nonumber\\
& &\times \prod_{j=1}^{l}H_{tot1}^{\gamma_j
v_{j},\gamma_{j+1}v_{j+1}} \delta_{\gamma_1\gamma}\delta_{v_{1}{v}}
\delta_{\gamma_{l+1}\gamma^\prime}\delta_{v_{l+1}v^\prime},\eeqa and
all $H_{tot1}^{\gamma_j
v_{\gamma_j},\gamma_{j+1}v_{j}}=\bra{\Phi^{\gamma_j
v_{j}}}H_{tot1}\ket{\Phi^{\gamma_{j+1}v_{j+1}}}$ form so-called
``perturbing Hamiltonian matrix", that is, the representation matrix
of the perturbing Hamiltonian in the unperturbed Hamiltonian
representation (SESR). While \beqa d_1(E[\gamma
v,l])&=&\prod_{i=1}^{l}\left(E_{\gamma_{1}v_1}
-E_{\gamma_{i+1}v_{j=1}}\right),\\
 d_i(E[\gamma v,l])&=&
\prod_{j=1}^{i-1}\left(E_{\gamma_{j}v_j}
-E_{\gamma_{i}v_i}\right)\!\!\!\prod_{k=i+1}^{l+1}\left(E_{\gamma_{i}v_i}
-E_{\gamma_{k}v_k}\right),\\[-3pt] d_{l+1}(E[\gamma v,l])
&=&\prod_{i=1}^{l}\left(E_{\gamma_{i}v_i}-E_{\gamma_{l+1}v_{l+1}}\right).\eeqa

By tracing off the degree of freedom of environment space, we obtain
the explicit expression of time evolution of reduced density matrix
of open system \beq\label{esos}
\rho_{\rm{S}}(t)=\sum_{k,l=0}^\infty\sum_{\beta, u,\beta^\prime,
u^\prime}\sum_{\gamma v,\gamma^\prime}A_k^{\gamma v,\beta
u}(t)\rho_{tot}^{\beta u,\beta^\prime u^\prime}(0)A_l^{\beta^\prime
u^\prime,\gamma^\prime
v}(-t)\ket{\phi^{\gamma}}\bra{\phi^{\gamma^\prime}},\eeq where we
have used the fact $\Tr_{\rm E}\left(\ket{{\Phi}^{\gamma
v}}\bra{{\Phi}^{\gamma^\prime v^\prime}}\right)
=\ket{{\psi}^{\gamma}}\bra{{\psi}^{\gamma^\prime}}\delta_{v
v^\prime}$, which is an advantage of the SESR.

It is clear that in the above expression, we need to know the
concrete forms of $\ket{\phi^\gamma}$ and $\ket{\chi^v}$ in order to
obtain the explicit expressions of $A_k^{\gamma v,\beta u}(t)$. In
fact, this is a physical reason why we take the form of $H_{tot0}$
as Eq.(\ref{H0form}) so that the eigenvectors and eigenvalues of
$H_{tot0}$ are obtainable.

Note that there are apparent divergences in the above exact
solution. For the tidiness in form, we keep these apparent
divergences in our expressions, but we can completely eliminate them
by the limit process \cite{My2}. In other words, our exact solution
of open systems should be understood in the limitation sense.

Just as pointed out above, there is, at least, an inherent SESR
(ISESR) in the total system if taking $H_{tot 0}=H_{\rm S}+H_{\rm
E}$. We will be able to obtain the similar solution as Eq.
(\ref{esos}). However, the ISESR is not unique in general because,
in principle, a part of $H_{\rm S}$ and/or a part of $H_{\rm E}$ can
be absorbed to $H_{\rm SE}$ if $I_{\rm S}$ and $I_{\rm E}$ are
thought of as, respectively, the system operator and the environment
operator. In this sense, the difference between the SESR and the
ISESR is that the SESR allows to contain a part of $H_{{\rm
SE}0}=\sum_{m,n} c_{mn}S_{m0}\otimes B_{n0}$, in which, $S_{m0}\neq
I_{\rm S}$ and $\sum_{n}c_{mn} B_{n0}\neq I_{\rm E}$ for all $m$. Of
course, if the cut-off approximation of perturbation is necessary,
it requires that the absorbed parts from $H_{\rm S}$ and $H_{\rm E}$
are small enough. Such an example is discussed in Sec. \ref{sec8}.
In addition, one of the reasons to introduce the SESR is to make the
Hamiltonian redivision and absorbing the perturbing parts of $H_{\rm
S}$ and $H_{\rm E}$ to the perturbing Hamiltonian of the total
system look like more natural.

Different from the general and explicit solution (\ref{gsrhos}), the
coefficients of our above solution (\ref{esos}) of open system
dynamics are $c$-number functions whose forms are expressed clearly.
Because $A_k^{\gamma v,\beta u}(t)$ include all of order
approximations, this solution is, in fact, exact although it is an
infinite series. Our solution in form is general enough, and it is
able to applied to the cases that $H_{\rm S}$ and/or $H_{\rm E}$ are
not exactly solvable. It is clear that we do not use the accustomed
approximations such as the Born-Markov approximation, the
factorization assumption for the initial state et. al. Hence, it
should be more general and more reliable in theory. Moreover, by
virtue of the improved scheme of perturbation theory proposed by us,
we can obtain the improved perturbed solution of open system
dynamics with better precision and higher efficiency because the
contributions from the high order even all order approximations of
perturbation are absorbed into the lower order approximations.

\section{Improved perturbed solution of open system
dynamics}\label{sec3}

Traditional scheme of perturbation theory has been successfully used
to solve many systems. However, in our point of view, it is still
improvable, even it has a flaw because it introduce the perturbing
parameter too early so that the contributions from the high order
even all order approximations of the diagonal and off-diagonal
elements of the perturbing Hamiltonian matrix are, respectively,
inappropriately dropped and prematurely cut off. For some systems,
the influences on the calculational precision because of this flaw
can be not neglectable with the evolution time increasing. Actually,
the traditional scheme of perturbation theory does not give a
general term form of expanding coefficient evolution with time for
any order approximation and does not explicitly express the general
term as an obvious $c$-number function. Thus, it is necessary to
find the perturbed solution (or perturbed energy and perturbed state
vector) from the low to the high order approximation step by step up
to some order approximation for a needed precision. Recently, in our
work, we proposed an improved scheme of perturbation theory based on
the general and explicit form of our exact solution \cite{My1,My2}.
In our improved scheme, we introduce the approximation as late as
possible, and consider subtly and systemically the affection of
high-order approximation to the low-order one by the dynamical
rearrangement and summation method. This finally results in the
improved form of perturbed solution, and its expansion coefficients
contain reasonably the high-order energy improvement. In this
section, we will apply our improved scheme of perturbation theory to
open systems.

It must be emphasized that before applying our improved form of
perturbed solution, we have to first carried out the digonalization
of degenerate subspaces if there is degeneracy and do the
Hamiltonian redivision when $H_{tot1}$ has the diagonal elements, in
order to completely removed possible degeneracies by this procedure.
When the remained degeneracies are allowed, it requires that the
off-diagonal elements of the perturbing Hamiltonian matrix between
any two degenerate levels are always vanishing. For more complicated
cases, we will study in the near further.

Therefore, up to the three order improved approximation, we have
\beqa\label{ipsos} \rho_{\rm{S}}(t)&=&\sum_{\stackrel{\scriptstyle
l,k=0}{k+l\leq 3}}^3\sum_{\beta, u,\beta^\prime,
u^\prime}\sum_{\gamma v,\gamma^\prime}A_{{\rm I}l}^{\gamma v,\beta
u}(t)\rho_{tot}^{\beta u,\beta^\prime u^\prime}(0)A_{{\rm
I}k}^{\beta^\prime u^\prime,\gamma^\prime
v}(-t)\ket{\phi^{\gamma}}\bra{\phi^{\gamma^\prime}}+\mathcal{O}\left(H_1^4\right),\eeqa
where
\beq A_{{\rm I}0}^{\gamma v, \gamma^\prime
v^\prime}(t)=\e^{-\I\widetilde{E}_{\gamma v} t}\delta_{\gamma
\gamma^\prime}\delta_{vv\prime},\eeq
\beqa A_{\rm I 1}^{\gamma v,\gamma^\prime
v^\prime}(t)&=&\left[\frac{\e^{-\I \widetilde{E}_{\gamma v}
t}}{E_{\gamma v}-E_{\gamma^\prime v^\prime}}-\frac{\e^{-\I
\widetilde{E}_{\gamma^\prime v^\prime} t}}{E_{\gamma
v}-E_{\gamma^\prime v^\prime}}\right]g_1^{\gamma v,\gamma^\prime
v^\prime},\eeqa
\beqa A_{\rm I2}^{\gamma v,\gamma^\prime
v^\prime}(t)&=&\sum_{\gamma_1,v1}\left\{-\frac{\e^{-\I
\widetilde{E}_{\gamma v} t}-\e^{-\I \widetilde{E}_{\gamma_1 v_1}
t}}{\left(E_{\gamma v}-E_{\gamma_1 v_1}\right)^2} g_1^{\gamma
v,\gamma_1 v_1}g_1^{\gamma_1 v_1, \gamma
v}\delta_{\gamma\gamma^\prime}
\delta_{vv^\prime}+\left[\frac{\e^{-\I \widetilde{E}_{\gamma v}
t}}{\left(E_{\gamma v}-E_{\gamma_1 v_1}\right)
\left(E_{\gamma v}-E_{\gamma^\prime v^\prime}\right)}\right.\right.\nonumber\\
& & \left.\left.-\frac{\e^{-\I \widetilde{E}_{\gamma_1 v_1}
t}}{\left(E_{\gamma v}-E_{\gamma_1 v_1}\right)\left(E_{\gamma_1
v_1}-E_{\gamma^\prime v^\prime}\right)} +\frac{\e^{-\I
\widetilde{E}_{\gamma^\prime v^\prime} t}}{\left(E_{\gamma
v}-E_{\gamma^\prime v^\prime}\right)\left(E_{\gamma_1
v_1}-E_{\gamma^\prime v^\prime}\right)}\right]g_1^{\gamma v,\gamma_1
v_1}g_1^{\gamma_1 v_1,\gamma^\prime v^\prime}\eta_{\gamma
v,\gamma^\prime v^\prime}\right\},\hskip 1.0cm\eeqa
\beqa A_{\rm I3}^{\gamma v,\gamma^\prime
v^\prime}(t)&=&\sum_{\gamma_1 v_1,\gamma_2 v_2}\left[-\frac{\e^{-\I
\widetilde{E}_{\gamma v} t}}{\left(E_{\gamma v}-E_{\gamma_1
v_1}\right)\left(E_{\gamma v}-E_{\gamma_2
v_2}\right)^2}-\frac{\e^{-\I \widetilde{E}_{\gamma v}
t}}{\left(E_{\gamma v}-E_{\gamma_1 v_1}\right)^2\left(E_{\gamma
v}-E_{\gamma_2 v_2}\right)}
\right.\nonumber\\
& &\left.+\frac{\e^{-\I \widetilde{E}_{\gamma_1 v_1}
t}}{\left(E_{\gamma v}-E_{\gamma_1 v_1}\right)^2\left(E_{\gamma_1
v_1}-E_{\gamma_2 v_2}\right)}-\frac{\e^{-\I \widetilde{E}_{\gamma_2
v_2} t}}{\left(E_{\gamma v}-E_{\gamma_2v_2}\right)^2
\left(E_{\gamma_1 v_1}-E_{\gamma_2 v_2}\right)}\right] g_1^{\gamma
v,\gamma_1 v_1}g_1^{\gamma_1 v_1\gamma_2 v_2}g_1^{\gamma_2
v_2,\gamma v}\delta_{\gamma\gamma^\prime}\delta_{vv^\prime}
\nonumber\\
& &-\sum_{\gamma_1}\left[\frac{\e^{-\I \widetilde{E}_{\gamma v}
t}}{\left(E_{\gamma v}-E_{\gamma_1 v_1}\right) \left(E_{\gamma
v}-E_{\gamma^\prime v^\prime}\right)^2}+\frac{\e^{-\I
\widetilde{E}_{\gamma v} t}}{\left(E_{\gamma v} -E_{\gamma_1
v_1}\right)^2\left(E_{\gamma v}-E_{\gamma^\prime
v^\prime}\right)}\right]g_1^{\gamma v,\gamma_1v_1}
g_1^{\gamma_1v_1,\gamma v} g_1^{\gamma v\gamma^\prime v^\prime}\nonumber\\
& &+\sum_{\gamma_1,\gamma_2}\left[\frac{\e^{-\I
\widetilde{E}_{\gamma v} t}\eta_{\gamma v,\gamma_2
v_2}}{\left(E_{\gamma v}-E_{\gamma_1v_1}\right) \left(E_{\gamma
v}-E_{\gamma_2 v_2}\right) \left(E_{\gamma v}-E_{\gamma^\prime
v^\prime}\right)}-\frac{\e^{-\I \widetilde{E}_{\gamma_1v_1}
t}\eta_{\gamma_1v_1,\gamma^\prime v^\prime}}{\left(E_{\gamma
v}-E_{\gamma_1v_1}\right)\left(E_{\gamma_1
v_1}-E_{\gamma_2v_2}\right)
\left(E_{\gamma_1 v_1}-E_{\gamma^\prime v^\prime}\right)}\right.\nonumber\\
& &\left.+\frac{\e^{-\I \widetilde{E}_{\gamma_2v_2} t}\eta_{\gamma
v,\gamma_2v_2}}{\left(E_{\gamma v}-E_{\gamma_2v_2}\right)
\left(E_{\gamma_1v_1}-E_{\gamma_2v_2}\right)
\left(E_{\gamma_2v_2}-E_{\gamma^\prime v^\prime}\right)}\right]
g_1^{\gamma v,\gamma_1v_1}
g_1^{\gamma_1v_1,\gamma_2v_2}g_1^{\gamma_2v_2,\gamma^\prime
v^\prime}\eta_{\gamma v,\gamma^\prime v^\prime},\eeqa where
$\delta_{\gamma\gamma^\prime}$ and $\delta_{vv^\prime}$ are the
usual discrete delta functions, while
$\eta_{\gamma\gamma^\prime}=1-\delta_{\gamma\gamma^\prime}$,
$\eta_{vv^\prime}=1-\delta_{vv^\prime}$, and $\eta_{\gamma
v,\gamma^\prime
v^\prime}=\eta_{\gamma\gamma^\prime}+\delta_{\gamma\gamma^\prime}\eta_{vv^\prime}
=\eta_{vv^\prime}+\eta_{\gamma\gamma^\prime}\delta_{vv^\prime}$.
Moreover, we have defined so-called improved form of perturbed
energy by\beq \label{iped} \widetilde{E}_{\gamma v}=E_{\gamma
v}+G_{\gamma v}^{(1)} +G_{\gamma v}^{(2)}+G_{\gamma
v}^{(3)}+G_{\gamma v}^{(4)}+G_{\gamma v}^{(5)}+\cdots ,\eeq where,
$G_{\gamma v}^{(1)}=h_1^{\gamma v}$ are diagonal elements of
$H_{tot1}$ and $g_1^{\gamma_i v_i,\gamma_j v_j}$ are off-diagonal
elements of $H_{tot1}$ in the representation of $H_{tot0}$. In
addition, $h_1^{\gamma v}$ include the diagonal elements after the
diagonalization of degenerate subspaces. While \beq G_{\gamma
v}^{(2)}=\sum_{\gamma_1,v_1}\frac{1}{E_{\gamma v}-E_{\gamma_1 v_1}}
g_{1}^{\gamma v, \gamma_1 v_1}g_{1}^{\gamma_1v_1,\gamma v}, \eeq
\beq G_{\gamma v}^{(3)}=\sum_{\gamma_1, v_1,\gamma_2, v_2}
\frac{1}{(E_{\gamma v}-E_{\gamma_1 v_1})(E_{\gamma v}-E_{\gamma_2
v_2})} g_{1}^{\gamma
v,\gamma_1v_1}g_{1}^{\gamma_1v_1,\gamma_2v_2}g_{1}^{\gamma_2v_2,\gamma
v}, \eeq \beqa G_{\gamma
v}^{(4)}&=&\sum_{\gamma_1,\gamma_2,\gamma_3}\sum_{v_1,v_2,v_3}
\frac{g_{1}^{\gamma v,\gamma_1v_1}g_{1}^{\gamma_1v_1,\gamma_2v_2}
g_{1}^{\gamma_2v_2,\gamma_3v_3}g_{1}^{\gamma_3v_3,\gamma v}
\eta_{\gamma v,\gamma_2v_2}}{(E_{\gamma v}-E_{\gamma_1 v_1})
(E_{\gamma v}-E_{\gamma_2 v_2})(E_{\gamma v}-E_{\gamma_3 v_3})}\nonumber\\
& &-\sum_{\gamma_1,\gamma_2}\sum_{v_1,v_2}\frac{g_{1}^{\gamma
v,\gamma_1v_1} g_{1}^{\gamma_1 v_1,\gamma v}g_{1}^{\gamma v,\gamma_2
v_2} g_{1}^{\gamma_2v_2,\gamma v}}{{(E_{\gamma v}-E_{\gamma_1
v_1})}^{2}(E_{\gamma v}-E_{\gamma_2v_2})}, \eeqa \beqa
G_\gamma^{(5)}&=&\sum_{\gamma_1,\gamma_2,\gamma_3,\gamma_4}
\sum_{v_1,v_2,v_3,v_4}\frac{g_1^{\gamma v,\gamma_1v_1}g_1^{\gamma_1
v_1,\gamma_2 v_2} g_1^{\gamma_2 v_2,\gamma_3 v_3}g_1^{\gamma_3
v_3,\gamma_4 v_4}g_1^{\gamma_4 v_4,\gamma v}\eta_{\gamma
v,\gamma_2v_2}\eta_{\gamma v,\gamma_3v_3}} {\left(E_{\gamma
v}-E_{\gamma_1 v_1}\right)\left(E_{\gamma v}-E_{\gamma_2 v_2}\right)
\left(E_{\gamma v}-E_{\gamma_3 v_3}\right)\left(E_{\gamma v}-E_{\gamma_4v_4}\right)}\nonumber\\
& & -\sum_{\gamma_1,\gamma_2,\gamma_3}\sum_{v_1,v_2,v_3}\left[
\frac{g_1^{\gamma v,\gamma_1v_1} g_1^{\gamma_1v_1,\gamma
v}g_1^{\gamma
v,\gamma_2v_2}g_1^{\gamma_2v_2,\gamma_3v_3}g_1^{\gamma_3v_3,\gamma
v}} {\left(E_{\gamma v}-E_{\gamma_1v_1}\right)^2\left(E_{\gamma
v}-E_{\gamma_2 v_2}\right) \left(E_{\gamma v}-E_{\gamma_3
v_3}\right)} \right.\nonumber\\
& &\left.+\frac{g_1^{\gamma v,\gamma_1v_1} g_1^{\gamma_1v_1,\gamma
v}g_1^{\gamma v,\gamma_2 v_2}g_1^{\gamma_2 v_2,\gamma_3
v_3}g_1^{\gamma_3 v_3,\gamma v}} {\left(E_{\gamma v}-E_{\gamma_1
v_1}\right)\left(E_{\gamma v}-E_{\gamma_2 v_2}\right)^2
\left(E_{\gamma v}-E_{\gamma_3 v_3}\right)}+\frac{g_1^{\gamma
v,\gamma_1 v_1} g_1^{\gamma_1 v_1,\gamma v}g_1^{\gamma
v,\gamma_2v_2}g_1^{\gamma_2v2,\gamma_3v_3}g_1^{\gamma_3v_3,\gamma
v}} {\left(E_{\gamma v}-E_{\gamma_1 v_1}\right)\left(E_{\gamma
v}-E_{\gamma_2 v_2}\right) \left(E_{\gamma v}-E_{\gamma_3
v_3}\right)^2}\right].\eeqa It must be emphasized that if only based
on our calculations that was completed in Ref. \cite{My2}, the
improved perturbed energy in the exponential powers of $A_{\rm I1}$,
$A_{\rm I2}$ and $A_{\rm I3}$ will be cut off, respectively, to
$G_{\gamma_iv_i}^{(4)}$, $G_{\gamma_iv_i}^{(3)}$ and
$G_{\gamma_iv_i}^{(2)}$. However, according to our conjecture, we
think that they can congruously written as the definition
(\ref{iped}).

Our improved perturbed solution inherits some features from our
exact solutions, for example, it is an explicit $c$-number function,
easy to calculate, does not need the extra approximations. In
principle, we should can calculate to any order of improved
approximation. It must be emphasized that our improved form of
perturbed solution absorbs the partial contributions from the high
order even all order approximations of perturbation. This means that
our solution has better precision and higher efficiency. In fact,
these advantages have been seen in our recent work \cite{My1,My2}.

\section{Master equation of open systems including all order approximations}\label{sec4}

Because we have obtained the general and explicit solution of the
open system dynamics when the Hamiltonians of the system, its
environment and the interaction between them are known, it is
unnecessary to derive out the dynamical equation of open systems.
However, in order to understand the affection from the environment,
compare our solution with the existed motion equations and reveal
the improvement of our method, we would like to discuss the motion
equation and master equation in this section.

It is more convenient to derive out the master equation in the
inherent SESR (ISESR) of open systems. That is, we take
$H_{tot}=H_{\rm S}+H_{\rm E}$. In fact, it make us more easily
compare our results with the existed ones. Obviously, the bases of
ISESR are $\ket{\psi^\gamma}\otimes\ket{\omega^v}$, that is \beqa
H_{\rm S}\ket{\psi^\gamma}\otimes\ket{\omega^v}
=E_{{\rm S}\gamma}\ket{\psi^\gamma}\otimes\ket{\omega^v},\\
 H_{\rm
E}\ket{\psi^\gamma}\otimes\ket{\omega^v}=\varepsilon_{{\rm E}
v}\ket{\psi^\gamma}\otimes\ket{\omega^v}.\eeqa Similar to the way in
Sec. \ref{sec2}, we can obtain the exact solutions $\rho_{tot}(t)$
and $\rho_{\rm S}(t)$. All we need to do is to change
$\ket{\phi^{\gamma}}\otimes \ket{\chi^v}$ as
$\ket{\psi^\gamma}\otimes\ket{\omega^v}$ and define the all matrix
elements in the ISESR, for example, ${A}_k^{\beta^\prime
u^\prime,\gamma^\prime
v^\prime}(t)=\bra{\psi^\gamma\omega^v}\mathcal{A}_k(t)
\ket{\psi^{\gamma^\prime}\omega^{v^\prime}}$. Therefore,
\beqa\label{estotisser}
 \rho_{tot}(t)&=&\sum_{k,l=0}^\infty\sum_{\beta, u,\beta^\prime,
u^\prime}\sum_{\gamma, v,\gamma^\prime, v^\prime}{A}_k^{\gamma
v,\beta u}(t)\varrho_{tot}^{\beta u,\beta^\prime
u^\prime}(0){A}_l^{\beta^\prime u^\prime,\gamma^\prime
v^\prime}(-t)\ket{\psi^{\gamma}}\bra{\psi^{\gamma^\prime}}\otimes
\ket{\omega^{v}}\bra{\omega^{v^\prime}}. \eeqa \beq\label{essisser}
 \rho_{\rm S}(t)=\sum_{\gamma v,\gamma^\prime}\sum_{k,l=0}^\infty\sum_{\beta, u,\beta^\prime,
u^\prime}{A}_k^{\gamma v,\beta u}(t)\varrho_{tot}^{\beta
u,\beta^\prime u^\prime}(0){A}_l^{\beta^\prime
u^\prime,\gamma^\prime
v}(-t)\ket{\psi^{\gamma}}\bra{\psi^{\gamma^\prime}}.\eeq

From the solution (\ref{estotisser}), it is easy to get that
$\Tr_{\rm E}\left\{\left[H_{\rm
S},{\rho}_{tot}(t)\right]\right\}=\left[h_{\rm S},\rho_{\rm
S}(t)\right]$ and $\Tr_{\rm E}\left\{\left[H_{\rm
E},{\rho}_{tot}(t)\right]\right\}=0$. Hence, \beq \label{des0}
\dot{\rho}_{\rm S}(t)=\Tr_{\rm E}\dot{\rho}_{tot}(t)=-\I\Tr_{\rm
E}\left\{\left[H_{tot},{\rho}_{tot}(t)\right]\right\}=-\I\left[h_{\rm
S},\rho_{\rm S}(t)\right]-\I\Tr_{\rm E}\left\{\left[H_{\rm
SE},{\rho}_{tot}(t)\right]\right\},\eeq where $h_{\rm S}=\Tr_{\rm
E}H_{\rm S}$. Denoting system operators by $S_m$ and bath operators
by $B_n^\prime$, the most general form of $H_{\rm SE}$ is \beq
\label{H1form} H_{tot1}=H_{\rm SE}=\sum_{m,n} c_{mn}S_{m}\otimes
B^\prime_n=\sum_{m}S_{m}\otimes B_m,\eeq where $B_m=\sum_n
c_{mn}B^\prime_n$. Substituting the above relation into Eq.
(\ref{des0}), we obtain the motion equation of open systems \beqa
\label{des1}\dot{\rho}_{\rm S}(t)&=&-\I\left[h_{\rm S},\rho_{\rm
S}(t)\right]-\I\sum_m \left[S_m,\Tr_{\rm E}\left\{\left(I_{\rm
S}\otimes B_m\right)\rho_{tot}(t)\right\}\right].\eeqa The second
term of its right side represents the influence of the environment
on the system.

In order to express the motion equation (\ref{des1}) of the open
systems in the explicit matrix form, we introduce so-called
factorizing initial state assumption, that is, the system and its
environment are uncorrelated initially such that the total density
matrix is a direct product of the system and its environment density
matrices, \beq \rho_{tot}(0)=\rho_{\rm{S}}(0)\otimes
\rho_{\rm{E}}(0).\eeq Its advantage is to make us easily consider
the actions of the operators on, respectively, the open system space
and its environment space, and finally we can easily trace off the
degree of freedom of environment space. In order to use this
advantage, we introduce two new operators \beqa
\mathcal{A}_{L}^{(k)}(t)&=&\mathcal{A}_k(t)\mathcal{A}^{-1}_0(t)
=\sum_{\beta,\beta^\prime}\mathcal{P}_{\rm
S}(\beta,\beta^\prime)\otimes
\mathcal{A}_{{\rm E}L}^{(k)}(t,\beta,\beta^\prime),\\
\mathcal{A}_{R}^{(l)}(-t)&=&\mathcal{A}^{-1}_0(-t)
\mathcal{A}_l(-t)=\sum_{\gamma,\gamma^\prime}\mathcal{P}_{\rm
S}(\gamma,\gamma^\prime)\otimes \mathcal{A}_{{\rm
E}R}^{(l)}(-t,\gamma,\gamma^\prime), \eeqa where $\mathcal{P}_{\rm
S}(\beta,\beta^\prime)=\ket{\psi^\beta}\bra{\psi^{\beta^\prime}}$
are the basis operators of the system Hilbert space
$\mathcal{H}_{\rm S}$, while the operators $\mathcal{A}_{{\rm
E}L}^{(k)}(t,\beta,\beta^\prime)$ and $\mathcal{A}_{{\rm
E}R}^{(k)}(-t,\gamma,\gamma^\prime)$ are defined in environment
Hilbert space $\mathcal{H}_{\rm E}$ by  \beqa \mathcal{A}_{{\rm
E}L}^{(k)}(t,\beta,\beta^\prime)&=&\sum_{u,u^\prime}{A}_k^{\beta
u,\beta^\prime u^\prime}(t)\e^{\I E_{\beta^\prime u^\prime}t}
\ket{\omega^u}\bra{\omega^{u^\prime}},\\
\mathcal{A}_{{\rm
E}R}^{(l)}(-t,\gamma,\gamma^\prime)&=&\sum_{v,v^\prime}\e^{-\I
E_{\gamma v}t}{A}_l^{\gamma v,\gamma^\prime v^\prime}(-t)
\ket{\omega^v}\bra{\omega^{v^\prime}}.\eeqa Thus, we see that
$\mathcal{A}_{L}^{(k)}(t)$ and $\mathcal{A}_{R}^{(l)}(-t)$ are
decomposed as the summations whose every terms with the form that
the open system parts and its environment parts are separate. Hence,
we obtain \beq\label{rhoinmatrix}
\rho_{tot}(t)=\sum_{\beta,\beta^\prime,\gamma,\gamma^\prime}\left[\mathcal{P}_{\rm
S}(\beta,\beta^\prime)\varrho_{\rm S}(t)\mathcal{P}_{\rm
S}(\gamma,\gamma^\prime)\right]\otimes\left[\mathcal{A}_{{\rm
E}L}^{(k)}(t,\beta,\beta^\prime)\varrho_{\rm E}(t)\mathcal{A}_{{\rm
E}R}^{(l)}(-t,\gamma,\gamma^\prime)\right],\eeq where \beqa
\varrho_{\rm
S}(t)&=&\e^{-\I h_{\rm S} t}\rho_{\rm S}(0)\e^{-\I h_{\rm S} t},\\
\varrho_{\rm E}(t)&=&\e^{-\I h_{\rm E} t}\rho_{\rm E}(0)\e^{-\I
h_{\rm E} t},\eeqa and then \beqa
\varrho_{tot}(t)=\mathcal{A}_0(t)\rho_{tot}(0)\mathcal{A}_0(-t)=\varrho_{\rm
S}(t)\otimes\varrho_{\rm E}(t).\eeqa

Substituting Eq. (\ref{rhoinmatrix}) into the motion equation
(\ref{des1}) it immediately follows that \beqa
\label{meem}\dot{\rho}_{\rm S}(t)&=&-\I\left[h_{\rm S},\rho_{\rm
S}(t)\right]-\I\sum_m\sum_{k,l=0}^\infty
\sum_{\beta,\beta^\prime,\gamma,\gamma^\prime}C^{m,kl}_{\beta\beta^\prime,\gamma\gamma^\prime}(t)
\left[S_m,\mathcal{P}_{\rm S}(\beta,\beta^\prime)\varrho_{\rm
S}(t)\mathcal{P}(\gamma,\gamma^\prime)\right],\eeqa where we have
used the fact that \beq
C^{m,kl}_{\beta\beta^\prime,\gamma\gamma^\prime}(t)=\Tr_{\rm
E}\left[B_m\mathcal{A}_{{\rm
E}L}^{(k)}(t,\beta,\beta^\prime)\varrho_{\rm E}(t)\mathcal{A}_{{\rm
E}R}^{(l)}(-t,\gamma,\gamma^\prime)\right].\eeq

Further deduction needs us to obtain the motion equation of
$\varrho_{\rm S}(t)$ that is expressed by $\rho_{\rm S}(t)$. In
fact, based on Eq. (\ref{rhoinmatrix}), we have \beqa \varrho_{\rm
S}(t)=\rho_{\rm S}(t)-\sum_{\stackrel{\scriptstyle
k,l=0}{k+l>0}}\sum_{\beta,\beta^\prime,\gamma,\gamma^\prime}
K^{kl}_{\beta\beta^\prime,\gamma\gamma^\prime}(t)
\left[\mathcal{P}_{\rm S}(\beta,\beta^\prime)\varrho_{\rm
S}(t)\mathcal{P}_{\rm S}(\gamma,\gamma^\prime)\right],\eeqa where we
define the coefficients \beq
K^{kl}_{\beta\beta^\prime,\gamma\gamma^\prime}(t)=\Tr_{\rm
E}\left[\mathcal{A}_{{\rm
E}L}^{(k)}(t,\beta,\beta^\prime)\varrho_{\rm E}(t)\mathcal{A}_{{\rm
E}R}^{(l)}(-t,\gamma,\gamma^\prime)\right].\eeq  Therefore, we can
use the iterative method to rewrite it as \beqa \label{iterative}
\varrho_{\rm S}(t)&=&\rho_{\rm S}(t)+\sum_{M=1}^\infty(-1)^M
\left[\prod_{m=1}^M\sum_{\stackrel{\scriptstyle
k_m,l_m=0}{k_m+l_m>0}}\sum_{\beta_m,\beta_m^\prime,\gamma_m,\gamma_m^\prime}
K^{k_ml_m}_{\beta_m\beta_m^\prime,\gamma_m\gamma_m^\prime}(t)\right]\nonumber\\
& &\times \left[\prod_{i=1}^M\mathcal{P}_{\rm
S}(\beta_i,\beta_i^\prime)\right] \rho_{\rm
S}(t)\left[\prod_{j=1}^M\mathcal{P}_{\rm
S}(\gamma_j,\gamma_j^\prime)\right].\eeqa Substituting it into Eq.
(\ref{meem}), we obtain \beqa\label{eme} \dot{\rho}_{\rm
S}(t)&=&-\I\left[h_{\rm S},\rho_{\rm
S}(t)\right]-\I\sum_m\sum_{k,l=0}^\infty
\sum_{\beta,\beta^\prime,\gamma,\gamma^\prime}C^{m,kl}_{\beta\beta^\prime,\gamma\gamma^\prime}(t)
\left[S_m,\mathcal{P}_{\rm S}(\beta,\beta^\prime)\rho_{\rm
S}(t)\mathcal{P}(\gamma,\gamma^\prime)\right]\nonumber\\
& &-\I\sum_m\sum_{k,l=0}^\infty
\sum_{\beta,\beta^\prime,\gamma,\gamma^\prime}C^{m,kl}_{\beta\beta^\prime,\gamma\gamma^\prime}(t)
\sum_{N=1}^\infty(-1)^N
\left(\prod_{n=1}^N\sum_{\stackrel{\scriptstyle
k_n,l_n=0}{k_n+l_n>0}}\sum_{\beta_n,\beta_n^\prime,\gamma_n,\gamma_n^\prime}
K^{k_nl_n}_{\beta_n\beta_n^\prime,\gamma_n\gamma_n^\prime}(t)\right)\nonumber\\
& &\times \left[S_m,\mathcal{P}_{\rm
S}(\beta,\beta^\prime)\left(\prod_{i=1}^N\mathcal{P}_{\rm
S}(\beta_i,\beta_i^\prime)\right) \rho_{\rm
S}(t)\left(\prod_{j=1}^N\mathcal{P}_{\rm
S}(\gamma_j,\gamma_j^\prime)\right)\mathcal{P}_{\rm
S}(\gamma,\gamma^\prime)\right]. \eeqa

Up to now, we have not introduced any approximation except for the
factorization assumption for the initial state. Since our master
equation (\ref{eme}) including all order approximations, we can say
it is an exact master equation of open systems.

\section{Perturbed master equation of open systems}\label{sec5}

In the most cases, the interaction between the open system and its
environment is weak. We can cut off the above exact master equation
to some given order approximation. It is clear that since we absorb
the coupling coefficients into $B_m$, we known
$C^{m,nl}_{\beta\beta^\prime,\gamma\gamma^\prime}(t)$ is a quantity
of the $(n+l+1)$th order approximation,
$K^{nl}_{\beta\beta^\prime,\gamma\gamma^\prime}(t)$ is a quantity of
the $(n+l)$th order approximation from their definitions. Although
we can obtain any given order approximation of master equation based
on our exact master equation (\ref{eme}), in most cases, we only are
interested in up to the second order approximation. Because \beq
C_{\beta\beta^\prime,\gamma\gamma^\prime}^{m,00}=\Tr_{\rm
E}\left[B_m\varrho_{\rm
E}(t)\right]\delta_{\beta\beta^\prime}\delta_{\gamma\gamma^\prime},\eeq
\beqa C_{\beta\beta^\prime,\gamma\gamma^\prime}^{m,0l}
&=&\delta_{\beta\beta^\prime}\Tr_{\rm E}\left[B_m\varrho_{\rm
E}(t)\mathcal{A}_{{\rm E}R}^{(l)}(-t,\gamma,\gamma^\prime)\right],\\
C_{\beta\beta^\prime,\gamma\gamma^\prime}^{m,k0} &=&\Tr_{\rm
E}\left[\mathcal{A}_{{\rm
E}L}^{(k)}(t,\beta,\beta^\prime)\varrho_{\rm E}(t)B_m\right]
\delta_{\gamma\gamma^\prime}.\eeqa
\beqa K_{\beta\beta^\prime,\gamma\gamma^\prime}^{0l}
&=&\delta_{\beta\beta^\prime}\Tr_{\rm E}\left[\varrho_{\rm
E}(t)\mathcal{A}_{{\rm E}R}^{(l)}(-t,\gamma,\gamma^\prime)\right],\\
K_{\beta\beta^\prime,\gamma\gamma^\prime}^{k0} &=&\Tr_{\rm
E}\left[\mathcal{A}_{{\rm
E}L}^{(k)}(t,\beta,\beta^\prime)\varrho_{\rm E}(t)\right]
\delta_{\gamma\gamma^\prime},\eeqa we have \beqa \label{meepf}
\dot{\rho}_{\rm S}(t)&=&-\I\left[h_{\rm S},\rho_{\rm
S}(t)\right]-\I[J(t),{\rho}_{\rm S}(t)]-\I\sum_m[S_m,{\rho}_{\rm
S}(t)C_{Rm}^{(1)}(t)+C_{Lm}^{(1)}(t){\rho}_{\rm S}(t)]\nonumber\\
& &+\I[J(t),{\rho}_{\rm S}(t)R^{(1)}(t)+L^{(1)}(t){\rho}_{\rm
S}(t)]+\mathcal{O}(H_{tot1}^3),\eeqa where \beq J(t)=\sum_m
{S}_m(t)\Tr_{\rm E}\left({B}_m\varrho_{\rm E}(t)\right),\eeq \beqa
C_{Lm}^{(1)}(t)&=&\Tr_{\rm E}\left\{\left[I_{\rm S}\otimes
B_m\right]\mathcal{A}_1(t)\e^{\I H_{tot0} t}\left[I_{\rm
S}\otimes\varrho_{\rm E}(t)\right]\right\},\\
C_{Rm}^{(1)}(t)&=&\Tr_{\rm E}\left\{\left[I_{\rm
S}\otimes\varrho_{\rm E}(t)\right]\e^{-\I H_{tot0}
t}\mathcal{A}_1(-t)\left[I_{\rm
S}\otimes B_m\right]\right\},\\
L^{(1)}(t)&=&\Tr_{\rm E}\left\{\mathcal{A}_1(t)\e^{\I H_{tot0}
t}\left[I_{\rm
S}\otimes\varrho_{\rm E}(t)\right]\right\},\\
R^{(1)}(t)&=&\Tr_{\rm E}\left\{\left[I_{\rm S}\otimes\varrho_{\rm
E}(t)\right]\e^{-\I H_{tot0} t}\mathcal{A}_1(-t)\right\},\eeqa while
\beqa \mathcal{A}_1(t)&=&\sum_{\beta,\gamma,u,v} \frac{\e^{-\I
E_{\beta u}t}-\e^{-\I E_{\gamma v}t}}{E_{\beta u}-E_{\gamma
v}}\left(\sum_m
S_m^{\beta\gamma}B_m^{uv}\right)\ket{\psi^\beta}\bra{\psi^\gamma}\otimes
\ket{\omega^u}\bra{\omega^v}\\
&=&\sum_{\beta,\gamma,u,v}\mathcal{A}_1^{\beta u,\gamma
v}(t)\ket{\psi^\beta}\bra{\psi^\gamma}\otimes
\ket{\omega^u}\bra{\omega^v}.\eeqa We can see that the Redfield
master equation will be obtained from our this master equation
without using Born-Markov approximation in next section.

In order to absorbing the partial contributions from the high order
even all order approximations into the lower order approximations,
we can use our improved scheme of perturbation theory. In similar
way used above, we have \beqa \label{meeI} \dot{\rho}_{\rm
S}(t)&=&-\I\left[h_{\rm S},\rho_{\rm
S}(t)\right]+\I\left[J(t),{\rho}_{\rm
S}(t)\right]-\I\sum_{a=0}^1\sum_m\left[S_m,{\rho}_{\rm
S}(t)C_{{\rm I}Rm}^{(a)}(t)+C_{{\rm I}Lm}^{(a)}(t){\rho}_{\rm S}(t)\right]\nonumber\\
& &-\I\left[J(t),{\rho}_{\rm S}(t)R_{\rm I}^{(1)}(t)+L_{\rm
I}^{(1)}(t){\rho}_{\rm S}(t)\right]+\I\sum_m\left[S_m,C_{{\rm
I}Lm}^{(0)}(t){\rho}_{\rm S}(t)R_{\rm I}^{(1)}(t)+C_{{\rm
I}Lm}^{(0)}(t)L_{\rm I}^{(1)}(t){\rho}_{\rm
S}(t)\right.\nonumber\\
& &\left.+{\rho}_{\rm S}(t)R_{\rm I}^{(1)}(t)C_{{\rm
I}Rm}^{(0)}(t)+L_{\rm I}^{(1)}(t){\rho}_{\rm S}(t)C_{{\rm
I}Rm}^{(0)}(t)\right] +\mathcal{O}(H_{tot1}^3),\eeqa where we have
defined \beqa C_{{\rm I}Lm}^{(k)}(t)&=&\Tr_{\rm
E}\left\{\left[I_{\rm S}\otimes B_m\right]\mathcal{A}_{{\rm
I}k}(t)\e^{\I H_{tot0} t}\left[I_{\rm
S}\otimes\varrho_{\rm E}(t)\right]\right\},\\
C_{{\rm I}Rm}^{(l)}(t)&=&\Tr_{\rm E}\left\{\left[I_{\rm
S}\otimes\varrho_{\rm E}(t)\right]\e^{-\I H_{tot0}
t}\mathcal{A}_{{\rm I}l}(-t)\left[I_{\rm
S}\otimes B_m\right]\right\},\\
L_{\rm I}^{(k)}(t)&=&\Tr_{\rm E}\left\{\mathcal{A}_{{\rm
I}k}(t)\e^{\I H_{tot0} t}\left[I_{\rm
S}\otimes\varrho_{\rm E}(t)\right]\right\},\\
R_{\rm I}^{(l)}(t)&=&\Tr_{\rm E}\left\{\left[I_{\rm
S}\otimes\varrho_{\rm E}(t)\right]\e^{-\I H_{tot0}
t}\mathcal{A}_{{\rm I}l}(-t)\right\},\eeqa while \beqa
\mathcal{A}_{\rm I1}(t)&=&\sum_{\beta,\gamma,u,v} \frac{\e^{-\I
\widetilde{E}_{\beta u}t}-\e^{-\I \widetilde{E}_{\gamma
v}t}}{E^\prime_{\beta u}-E^\prime_{\gamma
v}}H_{tot1}^{\prime}{}^{\beta u,\gamma
v}\left(1-\delta_{\beta\gamma}\delta_{uv}\right)
\ket{\psi^\beta}\bra{\psi^\gamma}\otimes
\ket{\omega^u}\bra{\omega^v}\\
&=&\sum_{\beta,\gamma,u,v}\mathcal{A}_{\rm I1}^{\beta u,\gamma
v}(t)\ket{\psi^\beta}\bra{\psi^\gamma}\otimes
\ket{\omega^u}\bra{\omega^v}.\eeqa Here, $\widetilde{E}_{\gamma
v}=E_{{\rm S}\gamma}+\varepsilon_{{\rm E} v}+h_{\gamma v}+G_{\gamma
v}^{(2)}+G_{\gamma v}^{(3)}+G_{\gamma v}^{(4)}+\cdots$,
$E^\prime_{\gamma v}=E_\gamma+\varepsilon_v+h_{\gamma v}$,
$h_{\gamma v}$ are diagonal elements of $H_{tot1}$, and the
perturbed part of Hamiltonian in $G^{(i)}_{\gamma v}$ has be
redivided as $H_{tot1}^\prime=H_{\rm SE}-\sum_{\gamma v}h_{\gamma
v}\ket{\psi^\gamma}\bra{\psi^\gamma}\otimes\ket{\omega^v}\bra{\omega^v}$,
that is, $g_1^{\gamma v, \gamma^\prime
v^\prime}=\bra{\psi^\gamma\omega^v}H_{tot1}^\prime\ket{\psi^\gamma\omega^v}$.

It must be emphasized that the operators in the above definitions
and expressions are defined in the ISESR (that has been diagoalized
in the degenerate subspaces if the degeneracy cases exist). However,
$A_{{\rm I}k}(\pm t)$ including $\widetilde{E}$ have to be
calculated using $H_1^\prime$ that is the perturbing Hamiltonian via
the redivision skill. Hence, it is important to distinguish
$H_{tot0}=H_{\rm S}+H_{\rm E}$, $H_{tot1}=H_{\rm SE}$ and their
redivision $H_{tot0}^\prime$, $H_{tot1}^\prime$ in spite of them in
the same ISESR. In addition, we assume all degeneracies are
completely removed by the diagonalization procedure of degenerate
subspaces and/or hamiltonian redivision for simplicity and
determination. If the remained degeneracies are allowed, it requires
that the off-diagonal elements of the perturbing Hamiltonian matrix
between any two degenerate levels are always vanishing, in order to
let our improved scheme of perturbation theory work well.

In the above derivation of our master equation, we do not use
Born-Markov approximation, but only standard cut-off approximation.
From our point of view, it is more reasonable in physics theory and
its precision and reliability should be better in practical
applications.

\section{Re-deduction of Redfield master equation}\label{sec6}

In order to compare our master equation with the known master
equations and illustrate the validness of our master equation, we
will deduce the Redfield master equation from our master equation
without using the Born-Markov approximation in this section. In
addition, we point out what differences between our master equation
and the existed one, and provide the comments on well-known
approximations using in open system dynamics.

Firstly, we assume a thermal equilibrium for the environment, that
is, \beq
 \rho_{\rm{E}}(0)=\frac{e^{-\beta_{\rm B} H_{\rm E}}}{\Tr e^{-\beta_{\rm B} H_{\rm E}}}
 =\frac{1}{Z}\sum_{v}e^{-\beta_{\rm B}\varepsilon_{{\rm
 E}v}}\ket{\omega^v}\bra{\omega^v}, \eeq
where $\beta_{\rm B}=1/k_{\rm B}T$ with $T$ the bath equilibrium.
This is justified when the environment is ``very large". Thus, it is
easy to get \beqa
L^{(1)}(t)&=&-R^{(1)}(t)=F^{(1)}(t)\nonumber\\
&=&\sum_{\beta,\gamma,u}\sum_{m}\frac{\e^{-\I\left(E_{{\rm
S}\beta}-E_{{\rm S}\gamma}\right)t}-1}{E_{{\rm S}\beta}-E_{{\rm
S}\gamma}}S_m^{\beta\gamma}B^{uu}\rho_{\rm
E}^u\ket{\psi^\beta}\bra{\psi^\gamma}\\
&=&-\I\e^{-\I h_{\rm S} t}\int_0^t\d \tau \Tr_{\rm
E}\left[\widehat{H}_{\rm SE}(\tau)\left(I_{\rm
S}\otimes\rho_E(0)\right)\right]\e^{\I h_{\rm S}
t}\\
&=&-\I\int_0^t\d \tau \Tr_{\rm E}\left[\overline{H}_{\rm
SE}(\tau)\left(I_{\rm S}\otimes\rho_E(0)\right)\right],\eeqa where
\beq \overline{H}_{\rm SE}(\tau)=\e^{-\I H_0 \tau}{H}_{\rm SE}\e^{\I
H_0 \tau}.\eeq

Likewise, we have \beqa C_{Lm}^{(1)}(t)&=&-\I\e^{-\I h_{\rm S}
t}\int_0^t\d \tau \Tr_{\rm E}\left[\left(I_{\rm
S}\otimes\widehat{B}_m(t)\right)\widehat{H}_{\rm
SE}(\tau)\left(I_{\rm S}\otimes\rho_E(0)\right)\right]\e^{\I h_{\rm
S}
t}\\
&=&-\I\int_0^t\d \tau \Tr_{\rm E}\left[\left(I_{\rm
S}\otimes{B}_m\right)\overline{H}_{\rm SE}(\tau)\left(I_{\rm
S}\otimes\rho_E(0)\right)\right],\eeqa \beqa
C_{Rm}^{(1)}(t)&=&\I\e^{-\I h_{\rm S} t}\int_0^t\d \tau \Tr_{\rm
E}\left[\left(I_{\rm S}\otimes\rho_E(0)\right)\widehat{H}_{\rm
SE}(\tau)\left(I_{\rm S}\otimes\widehat{B}_m(t)\right)\right]\e^{\I
h_{\rm S}
t}\\
&=&\I\int_0^t\d \tau \Tr_{\rm E}\left[\left(I_{\rm
S}\otimes{B}_m\right)\overline{H}_{\rm SE}(\tau)\left(I_{\rm
S}\otimes\rho_E(0)\right)\right].\eeqa

Therefore, our master equation (\ref{meepf}) up to the second order
approximation can be rewritten as \beqa \label{meepf1}
\dot{\rho}_{\rm S}(t)&=&-\I\left[h_{\rm S},\rho_{\rm
S}(t)\right]-\I[J(t),{\rho}_{\rm S}(t)]-\int_0^t\d \tau\Tr_{\rm
E}\left\{\left[H_{\rm SE},\left[\overline{H}_{\rm
SE}(\tau),{\rho}_{\rm
S}(t)\otimes\rho_{\rm E}(0)\right]\right]\right\}\nonumber\\
& &+\left[J(t),\int_0^t\d \tau\Tr_{\rm
E}\left\{\left[\overline{H}_{\rm SE}(\tau),{\rho}_{\rm
S}(t)\otimes\rho_{\rm E}(0)\right]\right]\right\}.\eeqa

If we introduce the interaction picture, that is, an operator
$\widehat{O}$ in this picture is defined by a corresponding operator
in the Schr\"odinger picture \beq \widehat{O}(t)=\e^{\I H_{tot0} t}O
\e^{-\I H_{tot0} t}.\eeq It is clear that for an operator $F_{\rm
S}$ in the open system Hilbert space and an operator $F_{\rm E}$ in
the environment Hilbert space, we have \beqa \widehat{F}_{\rm
S}(t)&=&\e^{\I h_{\rm S} t}F_{\rm S} \e^{-\I h_{\rm S} t},\\
\widehat{F}_{\rm E}(t)&=&\e^{\I h_{\rm E} t}F_{\rm E} \e^{-\I h_{\rm
E} t}.\eeqa It immediately follows the master equation in the
interaction picture: \beqa \label{meepf2}
\frac{\d\widetilde{\rho}_{\rm S}(t)}{\d
t}&=&-\I\left[\widetilde{J}(t),\widetilde{\rho}_{\rm
S}(t)\right]-\int_0^t\d \tau\Tr_{\rm
E}\left\{\left[\widetilde{H}_{\rm SE},\left[\widetilde{H}_{\rm
SE}(\tau),\widetilde{\rho}_{\rm
S}(t)\otimes{\rho}_{\rm E}(0)\right]\right]\right\}\nonumber\\
& &+\left[\widetilde{J}(t),\int_0^t\d \tau\Tr_{\rm
E}\left\{\left[\widetilde{H}_{\rm SE}(\tau),\widetilde{\rho}_{\rm
S}(t)\otimes{\rho}_{\rm E}(0)\right]\right]\right\}. \eeqa It must
be emphasized that $\widetilde{\rho}_{\rm S}(t)$ is equal to $\e^{\I
h_{\rm S} t}\rho_{\rm S}(t)\e^{-\I h_{\rm S} t}$, but not $\e^{\I
h_{\rm S} t}\rho_{\rm S}(0)\e^{-\I h_{\rm S} t}$.

In special,when we introduce the assumption \beq
\label{app1}\Tr_{\rm E}\left\{\left[\widetilde{H}_{\rm
SE},\rho_{tot}(0)\right]\right\}=0.\eeq we have\beq
\sum_m\sum_{\beta\beta^\prime,\gamma\gamma^\prime}
C_{\beta\beta^\prime,\gamma\gamma^\prime}^{m,00}\left[S_m,\mathcal{P}_{\rm
S}(\beta,\beta^\prime)\varrho_{\rm
S}(t)\mathcal{P}(\gamma,\gamma^\prime)\right]=\e^{-\I h_{\rm
S}t}\Tr_{\rm E}\left\{\left[\widetilde{H}_{\rm
SE},\rho_{tot}(0)\right]\right\}\e^{\I h_{\rm S}t}=0.\eeq Thus, Eq.
(\ref{meem}) becomes \beqa \label{meem1}\dot{\rho}_{\rm
S}(t)&=&-\I\left[h_{\rm S},\rho_{\rm
S}(t)\right]-\I\sum_m\sum_{\stackrel{\scriptstyle
k,l=0}{k+l>0}}^\infty
\sum_{\beta,\beta^\prime,\gamma,\gamma^\prime}C^{m,kl}_{\beta\beta^\prime,\gamma\gamma^\prime}(t)
\left[S_m,\mathcal{P}_{\rm S}(\beta,\beta^\prime)\varrho_{\rm
S}(t)\mathcal{P}(\gamma,\gamma^\prime)\right],\eeqa that is, the
perturbed form of master equation up to the second order
approximation reads \beq \label{meuseapp1}\dot{\rho}_{\rm
S}(t)=-\I\left[h_{\rm S},\rho_{\rm
S}\right]-\I\sum_m[S_m,{\rho}_{\rm
S}(t)C_{Rm}^{(1)}(t)+C_{Lm}^{(1)}(t){\rho}_{\rm S}(t)].\eeq This
means that the approximation ({\ref{app1}) leads to the following
terms \beq \label{app1d1}\I[J(t),{\rho}_{\rm
S}(t)]+\left[J(t),\int_0^t\d \tau\Tr_{\rm
E}\left\{\left[\overline{H}_{\rm SE}(\tau),{\rho}_{\rm
S}(t)\otimes\rho_{\rm E}(0)\right]\right]\right\},\eeq or,
equivalently, in the interaction picture \beq\label{app1d2}
-\I\left[\widetilde{J}(t),\widetilde{\rho}_{\rm
S}(t)\right]+\left[\widetilde{J}(t),\int_0^t\d \tau\Tr_{\rm
E}\left\{\left[\widetilde{H}_{\rm SE}(\tau),\widetilde{\rho}_{\rm
S}(t)\otimes{\rho}_{\rm E}(0)\right]\right]\right\}\eeq are dropped
by comparing Eq. (\ref{meuseapp1}) with Eq. (\ref{meepf1}) or
(\ref{meepf2}).

Usually, the approximation is thought of as a unimportant
restriction since one can absorb the the dropped terms into the
system Hamiltonian $H_{\rm S}$. However, based on the above result,
we think that the approximation (\ref{app1}) is a real assumption
because the second term in (\ref{app1d1}) or (\ref{app1d2}) is not
nontrivial and it can not be absorbed into $H_{\rm S}$ in general.
In other words, the the second order contribution to the master
equation from the second term in (\ref{app1d1}) or (\ref{app1d2})
should be considered and the approximation (\ref{app1}) should be
rechecked for the concrete open systems except for the cases when
$J(t)=0$. Actually, we think, if $J(t)$ is not equal to zero, the
last term appears in our master equation (\ref{meepf1}) or
(\ref{meepf2}) is obviously different from the existed master
equations.

It is very interesting, when the approximation (\ref{app1}) can be
used to some given open systems, we immediately from the equation
(\ref{meuseapp1}) obtain
\beqa\label{Redfieldme}\frac{\d\widetilde{\rho}_{\rm S}(t)}{\d
t}&=&-\int_0^t\d \tau\Tr_{\rm E}\left\{\left[\widetilde{H}_{\rm
SE},\left[\widetilde{H}_{\rm SE}(\tau),\widetilde{\rho}_{\rm
S}(t)\otimes{\rho}_{\rm E}(0)\right]\right]\right\}. \eeqa This is
just the well-known Redfield master equation. This conclusion
implies that the Redfield master equation is still valid without
introducing Born-Markov approximation. Therefore, we think that
Born-Markov approximation is unnecessary for the master equation
with the second order perturbed approximations. From our point of
view, this is a real physical reason why ones should use jointly
Born- and Markov approximations and why ones can obtain useful
conclusions in the cases without Born-Markov approximation. In fact,
those terms that are dropped by Born approximation are compensated
by Markov approximation. In other words, Born approximation plus
Markov approximation back to no approximation based on our results.

\section{Milburn dynamics for open systems}\label{sec7}

Historically, a useful dynamical model of open system is the Milburn
model \cite{Milburn}. It provides a way to describe so-called
``intrinsic" decoherence. However, in our point of view, perhaps, it
can be called external-external environment decoherence. That is,
Milburn dynamics might be alternatively explained as the effect of
environment of the composite system or the large environment of the
proper system. This explanation is, in fact, a conclusion from that
we believe the von Neumann equation is uniquely correct for a closed
system. One argues what mechanism results in that the external
influence is reflected by the extra term in Milburn dynamics. We can
not answer it at present, but we would like to ask what condition
changes the dynamics from the von Neumann's to the Milburn's. If the
answer is the Milburn dynamics is a nature of closed quantum
systems, then it is very difficult how to understand the free
parameter $\theta_0$.

Actually, only one can do something within the near environment in
order to control decoherence, for example, the self-interaction of
environment, and it is possible that one only knows how to
appropriately describe the dynamics of near environment and the
interaction between the system and near environment but are short of
the knowledge about the remote environment. Therefore, in this
section, we intend to use the Milburn model to consider the dynamics
of the composite system made up of the system and its near
environment. The conclusions obtained here imply the our solution
and methods are also applicable to more general open systems such as
the Milburn model.

Dynamics in the Milburn model replaces the usual von-Neumann
equation of the density matrix by
\beq\label{MEM}
\dot{\rho}_{tot}(t)=-\I[H_{tot},\rho_{tot}(t)]-\frac{\theta_0}{2}[H_{tot},[H_{tot},\rho_{tot}(t)]]
,\eeq
\noindent where $\theta_0$ is a constant meaning that there is some
minimum unitary-phase transformation. This implies that coherence is
destroyed as the physical properties of the system approach a
macroscopic level. Hence, seemingly, the ``intrinsic" decoherence
explanation looks like to be reasonable. However, the minimum
unitary-phase transformation is not clear. The parameter $\theta_0$
in the Milburn model is still ``free". In other words, $\theta_0$ is
not been given by the theory. If we think the extra term is resulted
in by the remote environment, $\theta_0$ should be able be known by
the experiment.

Now, we directly extend Milburn dynamics to a Milburn-type closed
quantum system consisting of the concerned system and its near
environment. The Hamiltonian in eq.(\ref{MEM}) still reads $
H_{tot}=H_{\rm S}+H_{\rm{E}}+H_{\rm{SE}}$. Here, a Milburn-type
closed quantum system is not really closed system from the view that
a really closed system must obey the von Neumann equation. Actually,
an alternative explanation is that a Milburn-type closed quantum
system is still affected by the remote (larger environment), and
this influence is represented by an extra term with $\theta_0$
multiplier because one cannot know the Hamiltonian of its remote
environment and the interaction form between the interesting system
and its remote environment. Obviously, when $\theta_0=0$, Milburn
dynamics back to von Neumann dynamics. This implies that the (very)
remote environment can be ignored.

The formal solution of Milburn dynamics for the composite system can
be written as \cite{Kimm} \beq \label{fs}
\rho_{tot}(t)=\exp\left\{-\I H_{tot}t-\theta_0 H_{tot}^2
t/2\right\}\left[\e^{\mathfrak{M}t}\rho_{tot}(0)\right]\exp\left\{\I
H_{tot}t-\theta_0 H_{tot}^2 t/2\right\}= \sum_k^\infty
M_k(t)\rho_{tot}(0)M_k^\dagger(t),\eeq
where ${\mathfrak{M}}$ is a superoperator, i.e,
${\mathfrak{M}}\rho_{tot}=\theta_0 H_{tot}\rho_{tot} H_{tot}$, and
the Kraus operators $M_k(t)$ is in the form \beq
M_k(t)=\sqrt{\frac{(\theta_0t)^k }{k!}}H_{tot}^k \exp\left\{-\I
H_{tot}t-\theta_0 H_{tot}^2 t/2\right\}. \eeq Without loss of
generality, using of the denotation $(A+B)^K=A^K+f^K(A,B)$ and
$f^0(A,B)=0$, we can write \beq M_k(t)=\sqrt{\frac{(\theta_0t)^k
}{k!}}\left[{H}_{tot0}^k \exp\left\{-\I {H}_{tot0}t-\theta_0
{H}_{tot0}^2 t/2\right\}+\sum_{n=0}^\infty\sum_{m=0}^n \frac{(-\I
t)^n}{n!}\left(\frac{\theta_0}{2\I}\right)^m
C_n^mf^{n+m+k}(H_{tot0},H_{tot1})\right]. \eeq

Just as we find the exact solution of open system in von Neumann
dynamics, we need a system-environment separated representation
(SESR), which has been given in Sec. \ref{sec2}. Thus, in this SESR,
based on the our expansion formula of operator binomials power
\cite{My1} we have \beqa \label{fk} f^K({H}_{tot0},H_{tot1})
&=&\sum_{l=1}^K\sum_{\gamma_1,\cdots,\gamma_{l+1}}\sum_{v_1,\cdots,v_{l+1}}
C_l^K({E}[\gamma v,l]) \left[\prod_{j=1}^lH_{tot1}^{\gamma_j
v_j,\gamma_{j+1},v_{j+1}}\right] \ket{\psi^{\gamma_1}
}\bra{\psi^{\gamma_{l+1}}}\otimes\ket{\omega^{v_1}}\bra{\omega^{v_{l+1}}}\label{fkc},\eeqa
\beq C^K_l({E}[\gamma v,l])=\sum_{i=1}^{l+1} (-1)^{i-1}
\frac{{E}_{\gamma v}^K}{d_i({E}[\gamma v,l])}, \eeq where
$d_i({E}[\gamma v,l])$ is defined in Sec. \ref{sec2}. Therefore, the
expression of $M_k(t)$ is changed to a summation according to the
order (or power) of the $H_{tot1}$ as follows \beqa \label{mk}
M_k^{\gamma v,\gamma^\prime v^\prime}(t)&=&
\sqrt{\frac{(\theta_0t)^k }{k!}}{E}_{\gamma v}^k g({E}_{\gamma
v},t)\delta_{\gamma\gamma^\prime}\delta_{v v^\prime} +
\sqrt{\frac{(\theta_0t)^k }{k!}}\sum_{l=1}^\infty
\sum_{\gamma_1,\cdots,\gamma_{l+1}}\sum_{v_1,\cdots,v_{l+1}}\left[
\sum_{i=1}^{l+1}(-1)^{i-1}\frac{{E}_{\gamma_i v_i}^k g({E}_{\gamma_i
v_i};t)}{d_i({E}[\gamma v,l])}\right]\nonumber\\
& &\times \prod_{j=1}^{l}H_{tot1}^{\gamma_j
v_j,\gamma_{j^\prime}v_j^\prime}
\delta_{\gamma_1\gamma}\delta_{\gamma_{l+1}\gamma^\prime}
\delta_{v_1v}\delta_{v_{l+1}v^\prime},\eeqa where the time evolution
function $g(x;t)$ with the exponential form is defined by \beq
g(x;t)=\exp\left\{-\I xt -\theta_0 x^2 t/2\right\}. \eeq Obviously,
$M_k^\dagger{}^{\gamma v,\gamma^\prime v^\prime}(t)$ can be given
via replacing $g(x;t)$ by $g^*(x;t)$. Furthermore, we obtain the
expression of time evolution of reduced density matrix of open
systems, that is a general and explicit solution of open systems in
Milburn dynamics: \beqa \label{se1oa}
\rho_{\mathcal{S}}(t)&=&\sum_{\beta,u,\beta^\prime,u^\prime}
\sum_{\gamma,v,\gamma^\prime,v^\prime}M_k^{\beta u,\beta
u^\prime}(t) \rho^{\beta^\prime u^\prime, \gamma^\prime v^\prime}(0)
M_k^\dagger{}^{\gamma^\prime v^\prime,\gamma v}\delta_{uv}\ket{\psi^\beta}\bra{\psi^\gamma}\nonumber\\
&=& \sum_{\beta,\gamma,v}\rho^{\beta v,\gamma v}(0)g({E}_{\beta
v}-{E}_{\gamma v};t) \ket{\psi^\beta}\bra{\psi^{\gamma}}+
\sum_{\beta,u}\sum_{\gamma^\prime,v^\prime,\gamma}\rho^{\beta
u,\gamma^\prime v^\prime}(0)
\sum_{l=1}^\infty\sum_{\gamma_1,\cdots,\gamma_{l+1}}\left[\sum_{i=1}^{l+1}(-1)^{i-1}\frac{
g({E}_{\beta u}-{E}_{\gamma_i v_i};t)}{d_k({E}[\gamma
v,l])}\right]\nonumber\\
& &\times
\left[\prod_{j=1}^{l}H_{tot1}^{\gamma_j\gamma_{j+1}}\right]
\delta_{\gamma^\prime \gamma_1}\delta_{v^\prime
v_1}\delta_{\gamma_{l+1} \gamma}\delta_{v_{l+1} u}
\ket{{\psi}^\beta}\bra{\psi^{\gamma}}+
\sum_{\beta,\beta^\prime,u^\prime}\sum_{\gamma,v}\rho^{\beta^\prime
u^\prime,\gamma v}(0) \nonumber\\
& & \times
\sum_{l=1}^\infty\sum_{\beta_1,\cdots,\beta_{l+1}}\left[\sum_{i=1}^{l+1}(-1)^{i-1}\frac{
g({E}_{\beta_i v_i}-{E}_{\gamma v};t)}{d_k({E}[\beta
u,l])}\right]\left[\prod_{j=1}^{l}H_{tot1}^{\beta_j\beta_{j+1}}\right]
\delta_{\beta \beta_1}\delta_{v,u_1}\delta_{\beta_{l+1}
\beta^\prime}\delta_{u_{l+1} u^\prime}
\ket{\psi^\beta}\bra{\psi^{\gamma}}\nonumber\\
& &+
\sum_{\beta,u,\beta^\prime,u^\prime}\sum_{\gamma,v,\gamma^\prime,v^\prime}\rho^{\beta^\prime
u^\prime,\gamma^\prime v^\prime}(0)\sum_{k=1}^\infty
\sum_{l=1}^\infty\sum_{\beta_1,\cdots,\beta_{k+1}}\sum_{\gamma_1,\cdots,\gamma_{l+1}}
\left[\sum_{i=1}^{k+1}\sum_{j=1}^{l+1}(-1)^{i+j}\frac{
g({E}_{\beta_i u_i}-{E}_{\gamma_j v_j};t)}
{d_i({E}[\beta u])d_j({E}[\gamma v,l])}\right]\nonumber\\
&
&\times\left[\prod_{i=1}^{k}H_{tot1}^{\beta_iv_i,\beta_{i+1}v_{i+1}}\right]
\left[\prod_{j=1}^{l}H_{tot1}^{\gamma_j
v_j,\gamma_{j+1}v_{j+1}}\right]
\delta_{\beta\beta_1}\delta_{\beta_{k+1}\beta^\prime}\delta_{uu_1}\delta_{u_{l+1}u^\prime}
\delta_{\gamma^\prime\gamma_1}\delta_{\gamma_{l+1}\gamma}\delta_{v^\prime
v_1}\delta_{v_{l+1}v}\delta_{uv}
\ket{\psi^{\beta}}\bra{{\psi}^{\gamma}}.\hskip 0.5cm \eeqa It is
clear that if $\theta_0=0$, this solution is just the form of
solution of van Neumann dynamics that is obtained in Sec.
\ref{sec2}. Usually, the finite (even often low) order approximation
about $H_{tot1}$ can be taken, thus this expression will be cut off
to the finite terms. Similar to the methods used in Secs.
\ref{sec3}, \ref{sec4} and \ref{sec5}, we can study the perturbed
solution and motion equation of open systems in Milburn dynamics. It
is not difficult, so we omit them in order to save space.

\section{Example and application}\label{sec8}

In order to concretely illustrate our general and explicit solution
of open system dynamics, we recall an exactly solvable two-state
open system for decoherence that was first introduced by Zurek
\cite{Zurek,Zurek1982}. In this Zurek model, the ``free"
(unperturbed) Hamiltonian $H_{\rm S0}$ and the self-interaction
(perturbing) $H_{\rm S1}$ of concerning two-state system and the
`free" (unperturbed) Hamiltonian $H_{\rm E0}$ and the
self-interaction (perturbing) $H_{\rm E1}$ of the environment are
taken as to be equal to zero. The total Hamiltonian of the composite
system made of the interesting system plus the environment only has
their interaction term, that is \beq \label{zmh} H_{\rm
Zurek}=H_{\rm SE}=\sigma^z_{\rm S}\otimes B_{z\rm E},\eeq where the
environment operator $B_{z\rm E}$ is defined by \beq B_{z\rm E}=
\sum_{k=1}^{N_{\rm E}} \left(\bigotimes_{i=1}^{k-1}I_{{\rm
E}i}\right)\otimes\left(Z_k\sigma^{z_k}\right)\otimes\left(\bigotimes_{j=k+1}^{N_{\rm
E}}I_{{\rm E}j}\right),\eeq and $N_{\rm E}$ is the degree of freedom
of the environment, which is very large even infinite.

It is clear that this Zurek model can be exactly solved out. Its
eigenvectors are so-called natural bases \beq\label{zms} \ket{n_{\rm
S}n_{\rm E}}=\ket{n_{\rm S};n_1,n_2\cdots}=\ket{n_{\rm
S}}\otimes\bigotimes_{k=1}^{N_{\rm E}}\ket{n_k},\eeq \vskip -0.5cm
where $n_{\rm S},n_1,n_2,\cdots=0,1$ and \vskip -0.5cm\beqa
\ket{0}_{\rm S}&=&\left(\begin{array}{c} 1\\0
\end{array}\right),\quad\ket{1}_{\rm S}=\left(\begin{array}{c} 0\\1
\end{array}\right),\\
\ket{0}_{k}&=&\left(\begin{array}{c} 1\\0
\end{array}\right),\quad\ket{1}_{k}=\left(\begin{array}{c} 0\\1 \end{array}\right)
.\eeqa The corresponding eigenvalues are \beq \label{zme}E_{n_{\rm
S}n_{\rm E}}=E_{n_{\rm S}n_1n_2\cdots}=(-1)^{n_{\rm
S}}\sum_{k=1}^{N_{\rm E}}(-1)^{n_k}Z_k.\eeq Note that we use a
simple notation $n_{\rm E}$ to denote $n_1,n_2,\cdots$ here and
after.

Now, let we solve this Zurek model by using our exact solution or
the improved form of the perturbed solution. From our point of view,
the assumption that $H_{\rm S}$ and $H_{\rm E}$ are taken as zero is
a theoretical simplification. In fact, we can think that $H_{\rm S}$
and $H_{\rm E}$ are constants so that we can absorb them into energy
eigenvalues or, equivalently, directly omit them since these
constants do not affect physics. Therefore, the base of the SESR can
be taken as the natural bases (\ref{zms}).

Since $H_{\rm SE}$ is completely diagonal in this SESR, that is \beq
\bra{m_{\rm S} m_{\rm E}}H_{\rm SE}\ket{n_{\rm S}n_{\rm
E}}=(-1)^{n_{\rm S}}\sum_{k=1}^{N_{\rm E}}(-1)^{n_k}Z_k
\delta_{m_{\rm S}n_{\rm S}}\prod_{i=1}^{N_{\rm
E}}\delta_{m_in_i}.\eeq We should use the Hamiltonian redivision
skill, and then \beq H_{tot0}^\prime=H_{Zurek}.\eeq It is easy to
get \beqa \widetilde{E}_{n_{\rm S}n_{\rm E}}&=&\widetilde{E}_{n_{\rm
S}n_1n2\cdots}=(-1)^{n_{\rm S}}\sum_{k=1}^{N_{\rm
E}}(-1)^{n_k}Z_k.\\
A_{{\rm I}l}(t)&=&0, \quad (l>0).\eeqa where we have used the fact
$g_1^{m_{\rm S}m_{\rm E},n_{\rm S}n_{\rm E}}=0$ based on
$H_{tot1}^\prime=0$. This means that the perturbed solution part of
higher than the zeroth order approximation is vanishing. Therefore,
our exact solution or the improved form of the perturbed solution
including only non-vanishing zeroth order part becomes \beqa
\rho_{Zurek}(t)&=&\sum_{m_{\rm S},n_{\rm S}=0}^1\sum_{m_{\rm
E},n_{\rm E}}\e^{-\I \widetilde{E}_{m_{\rm S}m_{\rm
E}}t}\rho^{m_{\rm S}m_{\rm E},n_{\rm S}n_{\rm E}}(0)\e^{\I
\widetilde{E}_{n_{\rm S}n_{\rm E}}t}\ket{m_{\rm S}m_{\rm
E}}\bra{n_{\rm S}n_{\rm E}}=\e^{-\I H_{Zurek}t}\rho(0)\e^{\I
H_{Zurek}t}.\eeqa Obviously, it is equal to the exact solution of
the Zurek model (\ref{zmh}) via directly solving it. Of course, the
solutions $\rho_{\rm S}(t)$ of this open system obtained by our
exact solution or improved form of perturbed solution formula or
directly solving method are consistent. Therefore, we can say our
improved form of perturbed solution indeed absorbs the contributions
from all order approximations of the perturbing Hamiltonian $H_{\rm
SE}$ since it is diagonal. In addition, we would like to point out
that although there are the degeneracies in $\widetilde{E}_{n_{\rm
S}n_{\rm E}}$ when $m_{\rm S}+m_k=n_{\rm S}+n_k$, our improved form
of perturbed solution can work well since $H_{tot1}^\prime=0$.

In order to reveal the advantages of our exact solution and
perturbed solution, we add two transverse fields, respectively, in
the system and the environment, that is \beqa
\label{ezm}H_{tot}&=&\mu\sigma^x_{\rm S}\otimes I_{\rm E}+H_{\rm
Zurek}+\sigma^z_{\rm S}\otimes B_{x\rm E}= \mu\sigma^x_{\rm
S}\otimes I_{\rm E}+\sigma^z_{\rm S}\otimes\left(B_{x\rm E}+B_{z\rm
E}\right),\eeqa where \beq B_{x\rm E}=\left[\sum_{k}^{N_{\rm
E}}\left(\bigotimes_{i=1}^{k-1}I_{{\rm
E}i}\right)\otimes\left(X_k\sigma^{x_k}_{\rm
E}\right)\otimes\left(\bigotimes_{j=k+1}^{N_{\rm E}}I_{{\rm
E}j}\right)\right].\eeq The problem only with the system transverse
field was studied in Ref. \cite{Cucchietti}. The model (\ref{ezm})
is not exactly solvable unless $\mu=0$. Obviously, there are four
kinds of the SESRs.

{\em Case one}: The Hamiltonian split is \beqa \label{hs0}
H_{tot0}=H_{\rm S}+H_{\rm E}=\mu\sigma^x_{\rm S}\otimes I_{\rm
E},\quad H_{tot1}=\sigma^z_{\rm S}\otimes\left(B_{x\rm E}+B_{z\rm
E}\right).\eeqa The bases of unperturbed SESR are \beq
\label{isser1}\ket{\psi_{\rm S}^{n_{\rm S}}\chi^{n_{\rm
E}}}=\ket{\psi_{\rm S}^{n_{\rm S}}}\otimes\bigotimes_{k=1}^{N_{\rm
E}}\ket{\chi^{n_k}},\eeq \vskip -0.5cm where \vskip -0.5cm\beqa
\ket{\psi^{n_{\rm S}}}&=&\frac{1}{\sqrt{2}}\left[\ket{0}_{\rm
S}+(-1)^{n_{\rm
S}}\ket{1}_{\rm S}\right]\\
\ket{\chi^{n_k}}&=&\frac{1}{\sqrt{X_k^2+\left(Z_k+(-1)^{n_k}Y_k\right)^2}}
\left[\left(Z_k+(-1)^{n_k}Y_k\right)\ket{0}_k+X_k\ket{1}_k\right].\eeqa
where $Y_k=\sqrt{X_k^2+Z_k^2}$. Here, $\ket{\chi^{n_k}}$ are the
eigenvectors of the environment operator $B_{x\rm E}+B_{z\rm
E}=\left(X_k\sigma^x+Z_k\sigma^z\right)$, and corresponding
eigenvalues are $(-1)^{n_k}Y_k$. Thus, the eigenvalues of $H_{tot0}$
acting on $\ket{\psi_{\rm S}^{n_{\rm S}}\chi^{n_{\rm E}}}$ are \beq
\label{h0e0} E_{n_{\rm S}n_{\rm E}}=E_{n_{\rm
S},n_1n_2\cdots}=\mu(-1)^{n_{\rm S}}.\eeq

{\em Case two}: The Hamiltonian split is the same as (\ref{hs0}),
and the corresponding eigenvalues of $H_{tot0}$ are then the same as
(\ref{h0e0}). But the bases of unperturbed SESR can be taken as \beq
\label{isser2}\ket{\psi^{n_{\rm S}}n_{\rm E}}=\ket{\psi^{n_{\rm
S}}}\otimes\bigotimes_{k=1}^{N_{\rm E}}\ket{n_k}.\eeq

{\em Case three}: The Hamiltonian split is \beqa \label{hs1}
H_{tot0}= \mbox{0 or constant}, \quad H_{tot1}=\mu\sigma^x_{\rm
S}\otimes I_{\rm E}+\sigma^z_{\rm S}\otimes\left(B_{x\rm E}+B_{z\rm
E}\right).\eeqa The bases of a selected SESR are just the natural
bases $\ket{n_{\rm S}n_{\rm E}}$ defined in (\ref{zms}). Then, we
use our Hamiltonian redivision skill to obtain \beq
\label{hsp1}H_{tot0}^\prime=H_{\rm Zurek},\quad
H_{tot1}^\prime=\mu\sigma^x_{\rm S}\otimes I_{\rm E}+\sigma^z_{\rm
S}\otimes B_{x\rm E}.\eeq The corresponding eigenvalues of
$H_{tot0}^\prime$ is given by (\ref{zme}).

{\em Case four}: The Hamiltonian split is the same as (\ref{hs1}).
But the bases of the unperturbed SESR can be chosen as \beq
\ket{n_{\rm S}\chi_{\rm n_{\rm E}}}=\ket{n_{\rm
S}}\otimes\bigotimes_{k=1}^{N_{\rm E}}\ket{\chi^{n_k}}.\eeq Then, we
use our Hamiltonian redivision skill to obtain \beq\label{hs2}
H_{tot0}^\prime=H_{\rm Zurek}+\sigma^z_{\rm S}\otimes B_{x\rm
E},\quad H_{tot1}^\prime=\mu\sigma^x_{\rm S}\otimes I_{\rm E}.\eeq
The corresponding eigenvalue is \beq \label{case4e0}
E^\prime_{n_{\rm S}n_{\rm E}}=E^\prime_{n_{\rm
S}n_1n_2\cdots}=(-1)^{n_{\rm S}}\sum_{k=1}^{N_{\rm
E}}(-1)^{n_k}Y_k.\eeq

It must be emphasized that the four kinds of choices on the SESRS
aim at the different preconditions if the cut-off approximation of
perturbation is necessary. Cases one and two are used to the
preconditions that $\mu\gg Z_k$ and/or $\mu\gg X_k$, that is, the
transverse field $\mu$ is strong. Case three is chosen when $Z_k\gg
\mu$ and $Z_k\gg X_k$. In other words, two transverse fields are
weak. Case four is suitable to solve the problem under $Z_k\gg \mu$
and/or $X_k\gg \mu$. This means that the transverse field $\mu$ is
weak.

It is easy to see that in cases one and two there are two degenerate
subspaces with $N_{\rm E}$ dimensions, which cannot be completely
removed via the usual diagonalization procedure of the degenerate
subspaces and our Hamiltonian redivision. However, the conditions
that degeneracies happen are $\delta_{m_{\rm S}n_{\rm s}}$. For case
one \beq g_1^{m_{\rm S}m_{\rm E},n_{\rm S}n_{\rm E}}=\delta_{m_{\rm
S}\left(1-n_{\rm S}\right)}\left[\sum_{k=1}^{N_{\rm
E}}(-1)^{m_k}Y_k\right]\prod_{l=1}^{N_{\rm E}}\delta_{m_ln_l},\eeq
while for case two \beq g_1^{m_{\rm S}m_{\rm E},n_{\rm S}n_{\rm
E}}=\delta_{m_{\rm S}\left(1-n_{\rm
S}\right)}\left[\sum_{k=1}^{N_{\rm E}}(-1)^{m_k}Z_k
\left(\prod_{l=1}^{N_{\rm E}}
\delta_{m_ln_l}\right)+\sum_{k=1}^{N_{\rm
E}}\left(\prod_{i=1}^{k-1}\delta_{m_in_i}\right)
X_k\delta_{m_k\left(1-n_k\right)}\left(\prod_{j=k+1}^{N_{\rm
E}}\delta_{m_jn_j}\right)\right].\eeq This implies such a fact that
in both case one and case two $g_1^{m_{\rm S}m_{\rm E},n_{\rm
S}n_{\rm E}}$ are vanishing between any two degenerate levels, that
is $g_1^{m_{\rm S}m_{\rm E},n_{\rm S}n_{\rm E}}\delta_{m_{\rm
S}n_{\rm s}}=0$. Therefore, our improved scheme of perturbation
theory can work well. However, note that the preconditions that
$\mu\gg Z_k$ and/or $\mu\gg X_k$ in cases one and two are the same,
we prefer to use the choice of case one because its calculation is
easier than case two in our improved scheme of perturbation theory.

As to case three, we also can not completely remove the degeneracies
via the usual diagonalization procedure of the degenerate subspaces
and our Hamiltonian redivision. The conditions that degeneracies
happen are solutions of the following equation\beq
\sum_{k=1}^{N_{\rm E}}Z_k\left[(-1)^{m_{\rm S}+m_k}-(-1)^{n_{\rm
S}+n_k}\right]=0,\eeq while the off-diagonal elements of the
perturbing Hamiltonian matrix are \beq g_1^{m_{\rm S}m_{\rm
E},n_{\rm S}n_{\rm E}}=\mu\delta_{m_{\rm S}\left(1-n_{\rm
S}\right)}\prod_{k=1}^{N_{\rm E}}\delta_{m_k,n_k}+\delta_{m_{\rm
S}n_{\rm S}}\sum_{k=1}^{N_{\rm
E}}\left(\prod_{i=1}^{k-1}\delta_{m_in_i}\right)
X_k\delta_{m_k\left(1-n_k\right)}\left(\prod_{j=k+1}^{N_{\rm
E}}\delta_{m_jn_j}\right).\eeq It is clear that we can not
guarantee, in general, that $g_1^{m_{\rm S}m_{\rm E},n_{\rm S}n_{\rm
E}}$ are vanishing between any two degenerate levels. If the result
is indeed so. This SESR is not a good choice because the remained
degeneracies will result in the difficulty to use the usual
perturbation theory and complication in our improved ones. If we
still intend to use the cut-off approximation of perturbation, the
results from case three will not be satisfied enough if the
evolution time is long enough, because $A_{l}(t)$ or $A_{{\rm
I}l}(t)$ has the extra terms proportional to the evolution time that
can not be simply absorbed to the exponential power for $l\geq 2$ in
our views.

Fortunately, that case four can be covered the precondition that
$Z_k\gg \mu$ and $Z_k\gg X_k$ in case three. Hence, we give up the
choice of case three and only use the SESR in case four. Actually,
the above problems originally motivate us to consider how to choose
the appropriate SESR for open systems, which has been seen in Sec.
\ref{sec2}.

It is easy to get that the conditions that degeneracies happen in
case four are solutions of the following equation\beq\label{case4dc}
\sum_{k=1}^{N_{\rm E}}Y_k\left[(-1)^{m_{\rm S}+m_k}-(-1)^{n_{\rm
S}+n_k}\right]=0,\eeq while the off-diagonal elements of the
perturbing Hamiltonian matrix are \beq \label{case4e1}g_1^{m_{\rm
S}m_{\rm E},n_{\rm S}n_{\rm E}}=\mu\delta_{m_{\rm S}\left(1-n_{\rm
S}\right)}\prod_{k=1}^{N_{\rm E}}\delta_{m_k,n_k}.\eeq Our
interesting task is to seek for the conditions that degeneracies
happen when $g_1^{m_{\rm S}m_{\rm E},n_{\rm S}n_{\rm E}}\neq 0$.
Hence, we substitute $m_k=n_k$ for any $k$, into Eq. (\ref{case4dc})
and rewrite it as\beq \left[\sum_{k=1}^{N_{\rm
E}}Y_k(-1)^{m_k}\right]\left[(-1)^{m_{\rm S}}-(-1)^{n_{\rm
S}}\right]=0.\eeq Its solution is $m_{\rm S}=n_{\rm S}$ unless the
exception $\sum_{k=1}^{N_{\rm E}}Y_k(-1)^{m_k}=0$. However, this
exception is not valid limiting our problem to open systems because
it means that $H_{tot0}^\prime=0$ from Eq. (\ref{case4e0}), or
equivalently, the total $H_{tot}$ becomes $\mu\sigma^x_{\rm
S}\otimes I_{\rm E}$. Again jointly considering it and the
expression (\ref{case4e1}) of $g_1^{m_{\rm S}m_{\rm E},n_{\rm
S}n_{\rm E}}$ in case four, we obtain the conclusion that
$g_1^{m_{\rm S}m_{\rm E},n_{\rm S}n_{\rm E}}$ are indeed vanishing
between any two degenerate levels.

In the following discussion, we only focus on case one with the
strong transverse field $\mu$ and case four with weak transverse
field $\mu$ in order to illustrate our exact solution and improved
form of perturbed solution more simply and better.

Let us define \beq \delta_{m_{\rm E}n_{\rm E}}=\prod_{i=1}^{N_{\rm
E}}\delta_{m_in_i}.\eeq \beq f_{n_{\rm E}}=\sum_{n_1,n_2,\cdots=0}^1
Y_k(-1)^{n_k}=\sum_{n_{\rm E}=0}^1 Y_k(-1)^{n_k}.\eeq  Thus, for
case one, we can rewrite the off-diagonal elements of the perturbing
Hamiltonian matrix as \beq g_1^{m_{\rm S}m_{\rm E},n_{\rm S}n_{\rm
E}}=\delta_{m_{\rm S}\left(1-n_{\rm S}\right)}f_{m_{\rm
E}}\delta_{m_{\rm E}n_{\rm E}}.\eeq Substituting it into the
definition of $G_{n_{\rm S}n_{\rm E}}^{(a)}$, we obtain \beq
G_{n_{\rm S}n_{\rm E}}^{(2)}=\frac{(-1)^{n_{\rm S}}f^2_{n_{\rm
E}}}{2\mu},\quad G_{n_{\rm S}n_{\rm E}}^{(3)}=0, \quad G_{n_{\rm
S}n_{\rm E}}^{(4)}=-\frac{(-1)^{n_{\rm S}}f^4_{n_{\rm E}}}{8\mu^3}
.\eeq Hence, we have \beq \widetilde{E}_{n_{\rm S}n_{\rm
E}}=(-1)^{n_{\rm S}}\mu\left[1+\frac{1}{2}\left(\frac{f_{n_{\rm
E}}}{2\mu}\right)^2-\frac{1}{2}\left(\frac{f_{n_{\rm
E}}}{2\mu}\right)^4\right].\eeq

Similarly, for case four, from Eqs. (\ref{case4e0}) and
(\ref{case4e1}) it follows that \beq G_{n_{\rm S}n_{\rm
E}}^{(2)}=\frac{(-1)^{n_{\rm S}}\mu^2}{2f_{n_{\rm E}}}, \quad
G_{n_{\rm S}n_{\rm E}}^{(3)}=0, \quad G_{n_{\rm S}n_{\rm
E}}^{(4)}=-\frac{(-1)^{n_{\rm S}}\mu^4}{8f^3_{n_{\rm E}}}.\eeq
Hence, we have \beq \label{case4ei} \widetilde{E}_{n_{\rm S}n_{\rm
E}}=(-1)^{n_{\rm S}}f_{n_{\rm
E}}\left[1+\frac{1}{2}\left(\frac{\mu}{2f_{n_{\rm
E}}}\right)^2-\frac{1}{2}\left(\frac{\mu}{2f_{n_{\rm
E}}}\right)^4\right].\eeq

It is clear that there is a corresponding relation between the case
of the strong transverse field $\mu$ and the case of weak transverse
field $\mu$, that is, their perturbed solutions will be the same
under the exchanging transformation $\mu\Leftrightarrow f_{n_{\rm
E}}$. Hence, we only write down, for case one, the zeroth, first and
second order parts of total system density matrix at time $t$,
respectively\beq \rho_{tot}^{(0)}(t)=\sum_{m_{\rm S},m_{\rm
E}=0}^1\sum_{n_{\rm S},n_{\rm
E}=0}^1\e^{-\I\left(\widetilde{E}_{m_{\rm S}m_{\rm
E}}-\widetilde{E}_{n_{\rm S}n_{\rm E}}\right)t}\rho_{tot}^{m_{\rm
S}m_{\rm E},n_{\rm S}n_{\rm E}}(0)\ket{\psi_{\rm S}^{m_{\rm
S}}}\bra{\psi_{\rm S}^{n_{\rm S}}}\otimes\ket{\chi^{m_{\rm
E}}}\bra{\chi^{n_{\rm E}}},\eeq \beqa
\rho_{tot}^{(1)}(t)&=&\sum_{m_{\rm S},m_{\rm E}=0}^1\sum_{n_{\rm
S},n_{\rm E}=0}^1\left(\e^{-\I\left(\widetilde{E}_{m_{\rm S}m_{\rm
E}}-\widetilde{E}_{n_{\rm S}n_{\rm
E}}\right)t}-\e^{-\I\left(\widetilde{E}_{m_{\rm S}m_{\rm
E}}-\widetilde{E}_{\left(1-n_{\rm S}\right)n_{\rm
E}}\right)t}\right)\frac{(-1)^{n_{\rm S}}f_{n_{\rm E}}}{2\mu}\nonumber\\
& &\times \rho_{tot}^{m_{\rm S}m_{\rm E},n_{\rm S}n_{\rm
E}}(0)\ket{\psi_{\rm S}^{m_{\rm S}}}\bra{\psi_{\rm S}^{1-n_{\rm
S}}}\otimes\ket{\chi^{m_{\rm E}}}\bra{\chi^{n_{\rm
E}}}\nonumber\\
& &+\sum_{m_{\rm S},m_{\rm E}=0}^1\sum_{n_{\rm S},n_{\rm
E}=0}^1\left(\e^{-\I\left(\widetilde{E}_{m_{\rm S}m_{\rm
E}}-\widetilde{E}_{n_{\rm S}n_{\rm
E}}\right)t}-\e^{-\I\left(\widetilde{E}_{\left(1-m_{\rm
S}\right)m_{\rm E}}-\widetilde{E}_{n_{\rm S}n_{\rm
E}}\right)t}\right)\frac{(-1)^{m_{\rm S}}f_{m_{\rm E}}}{2\mu}\nonumber\\
& &\times \rho_{tot}^{m_{\rm S}m_{\rm E},n_{\rm S}n_{\rm
E}}(0)\ket{\psi_{\rm S}^{1-m_{\rm S}}}\bra{\psi_{\rm S}^{n_{\rm
S}}}\otimes\ket{\chi^{m_{\rm E}}}\bra{\chi^{n_{\rm E}}},\eeqa \beqa
\rho_{tot}^{(2)}(t)&=&-\sum_{m_{\rm S},m_{\rm E}=0}^1\sum_{n_{\rm
S},n_{\rm E}=0}^1\left(\e^{-\I\left(\widetilde{E}_{m_{\rm S}m_{\rm
E}}-\widetilde{E}_{n_{\rm S}n_{\rm
E}}\right)t}-\e^{-\I\left(\widetilde{E}_{m_{\rm S}m_{\rm
E}}-\widetilde{E}_{\left(1-n_{\rm S}\right)n_{\rm
E}}\right)t}\right)\left(\frac{f_{n_{\rm E}}}{2\mu}\right)^2\nonumber\\
& &\times \rho_{tot}^{m_{\rm S}m_{\rm E},n_{\rm S}n_{\rm
E}}(0)\ket{\psi_{\rm S}^{m_{\rm S}}}\bra{\psi_{\rm S}^{n_{\rm
S}}}\otimes\ket{\chi^{m_{\rm E}}}\bra{\chi^{n_{\rm
E}}}\nonumber\\
& &-\sum_{m_{\rm S},m_{\rm E}=0}^1\sum_{n_{\rm S},n_{\rm
E}=0}^1\left(\e^{-\I\left(\widetilde{E}_{m_{\rm S}m_{\rm
E}}-\widetilde{E}_{n_{\rm S}n_{\rm
E}}\right)t}-\e^{-\I\left(\widetilde{E}_{\left(1-m_{\rm
S}\right)m_{\rm E}}-\widetilde{E}_{n_{\rm S}n_{\rm
E}}\right)t}\right)\left(\frac{f_{m_{\rm E}}}{2\mu}\right)^2\nonumber\\
& &\times \rho_{tot}^{m_{\rm S}m_{\rm E},n_{\rm S}n_{\rm
E}}(0)\ket{\psi_{\rm S}^{m_{\rm S}}}\bra{\psi_{\rm S}^{n_{\rm
S}}}\otimes\ket{\chi^{m_{\rm E}}}\bra{\chi^{n_{\rm
E}}}\nonumber\\
& &+\sum_{m_{\rm S},m_{\rm E}=0}^1\sum_{n_{\rm S},n_{\rm
E}=0}^1\left(\e^{-\I\widetilde{E}_{m_{\rm S}m_{\rm
E}}t}-\e^{-\I\widetilde{E}_{\left(1-m_{\rm S}\right)m_{\rm
E}}t}\right)\left(\e^{-\I\widetilde{E}_{n_{\rm S}n_{\rm
E}}t}-\e^{-\I\widetilde{E}_{\left(1-n_{\rm S}\right)n_{\rm
E}}t}\right)\nonumber \\
& &\times \frac{(-1)^{m_{\rm S}+n_{\rm S}}f_{m_{\rm E}}f_{n_{\rm
E}}}{4\mu^2}\rho_{tot}^{m_{\rm S}m_{\rm E},n_{\rm S}n_{\rm
E}}(0)\ket{\psi_{\rm S}^{1-m_{\rm S}}}\bra{\psi_{\rm S}^{1-n_{\rm
S}}}\otimes\ket{\chi^{m_{\rm E}}}\bra{\chi^{n_{\rm E}}}.\eeqa It is
easy to give the solution of the reduced density matrix of the open
system up to the improved form of the second order approximation by
tracing off the environment space, that is \beq \rho_{\rm
S}(t)=\Tr_{\rm
E}\left[\rho_{tot}^{(0)}(t)+\rho_{tot}^{(1)}(t)+\rho_{tot}^{(2)}(t)\right].\eeq
For a given initial state, this trace is very easy to calculate and
the explicit form of of the open system solution is obtained. Then
we can discuss the decoherence and entanglement dynamics according
to the methods in, for example, \cite{Cucchietti,Ourmd}, and they
are arranged in our forthcoming manuscript (in preparing)
\cite{Ournew}.

\section{Discussion and conclusion}\label{sec9}

This paper studies open systems dynamics, which is the third in our
serial studies on quantum mechanics in general quantum systems. Its
conclusions are obtained based on our previous two works
\cite{My1,My2}.

It must be emphasized that we study open system dynamics according
to the ``the first principle", that is, the Schr\"odinger equation
and the von Neumann equation, and we do not consider the
phenomenological methods and theories. For generality in theory, we
obtain the exact solution of the open system without using any
approximation. The deduction of our exact master equation only uses
the factorizing initial condition. Particularly, we derive out our
perturbed master equation and its improved form, but we give up all
of approximations used in the traditional methods and formulism
except for the factorizing initial condition. It is very interesting
that we get the Redfield master equation without using the
Born-Markov approximation. This implies the Born-Markov
approximation is unnecessary based on our results.

A simple but key idea to obtain our exact solution of open systems
is an appropriate choice of the SESR. In fact, it closely connects
with the Hamiltonian redivision skill \cite{My2}. In Sec.
\ref{sec8}, we have clearly stated its reasons. Originally, the aim
that we propose this idea is to break the accustomed choice of
$H_{\rm S}+H_{\rm E}$, and build a picture to allow the interaction
between the open system and its environment into our unperturbed
representation. This makes the Hamiltonian redivision skill to look
like more natural.

Our exact solution and master equation of open systems are general
and explicit in form because all order approximations of the
perturbing Hamiltonian not only are completely included but also are
clearly expressed, although it is an infinite series. In special,
they are in $c$-number function forms rather than operator forms.
This means that they can inherit the same advantage as the Feynman
path integral expression. Moreover, they are power series of the
perturbing Hamiltonian like as the Dyson series in the interaction
picture. This implies that the cut-off approximation of perturbation
can be made for the needed precision of the problems.

Based on our improved scheme of perturbation theory, the improved
forms of perturbed solution and perturbed master equation can absorb
the partial contributions from the high order even all order
approximations of perturbation. Therefore, we can say that our open
system dynamics is actually calculable, operationally efficient,
conclusively more accurate.

In order to extend our method, we also discuss Milburn model of open
systems. In fact, from our point of view, Miburn model of dynamics
should be applied to so-called Milburn-type closed quantum systems
made up of the interesting open system and its near environment. A
Milburn-type closed quantum system is not really closed system from
the view that a really closed system must obey the von Neumann
equation. If one cannot know the Hamiltonian of its remote
environment and the interaction form between the interested system
and its remote environment, Milburn model of dynamics might be a
choice scheme to study this kind of open systems. In the above
sense, the extra term with $\theta_0$ multiplier in the Milburn
equation represents the influence from the remote environment.
Obviously, when $\theta_0=0$, Milburn dynamics back to von Neumann
dynamics. This implies that the (very) remote environment can be
ignored. We obtain the exact solution that can provide a general
tool to investigate those interesting and complicated open systems
when the environment model is partially known. However, there a free
parameter $\theta_0$ in the Milburn model. It is still not been
given by the theory, but it should be able be known by the
experiment if we think the extra term in the Milburn dynamics is
resulted in by the remote environment.

Note that our open system dynamics is derived from the first
principle, our open system dynamics is not applicable to the cases
that ones do not clearly know the Hamiltonians of the open system,
its environment and the interaction between the system and the
environment unless at this time Milburn model is suitable. How to
relate with some phenomenological theory of open systems will be
done in the near future.

As examples, Zurek model of two-state open system and its extension
with two transverse fields are studied, respectively, in the strong
and weak fields acting on the system. We specially display how to
choose the appropriate SESR. They indicate that our open system
dynamics is a powerful theory and tool. We are sure that our open
system dynamics can be used to more open systems since its
generality and clearness, and the calculations are simpler and more
efficient, the results are more accurate and more reliable than the
existed scheme.

In summary, our results can be thought of as theoretical
developments of open system dynamics, and they are helpful for
understanding the theory of quantum mechanics and providing some
powerful tools for the calculation of decoherence, entanglement
dynamics, quantum dissipation, quantum transport in general quantum
systems and so on. Together with our exact solution and perturbation
theory \cite{My1,My2}, they can finally form the foundation of
theoretical formulism of quantum mechanics in general quantum
systems. Further study on quantum mechanics of general quantum
systems is on progressing.

\section*{Acknowledgments}

We are grateful all the collaborators of our quantum theory group in
the institute for theoretical physics of our university. This work
was funded by the National Fundamental Research Program of China
under No. 2001CB309310, partially supported by the National Natural
Science Foundation of China under Grant No. 60573008.


\vskip -0.1in
\begin{references}
\bibitem{Carmichael}H. Carmichael, {\em An Open System Approach to Quantum Optics},
(Berlin, Heidelberg, 1994):
\bibitem{Plenio}M. B Plenio and P. L. Knight, Rev. Mod. Phys. {\bf
70}, 101 (1998)
\bibitem{Nielson}M. A, Neilson and I. C. Chuang, {\em Quantum
computation and quantum information}, Cambridge University Press
(2000)
\bibitem{Perskill}J. Perskill, {\em Physics 229: Advanced Mathematical
Methods of Physics -- Quantum computation and Information},
Califoria Institute of Technology, 1998. URL:
http://www.theory.caltech.edu/people/perskill/ph229/
\bibitem{Zurek}W. H. Zurek, Rev. Mod. Phys. {\bf 75}, 715(2003)
\bibitem{Schlosshauer}M. Schlosshauer, Rev. Mod. Phys. {\bf 76},
1267(2004)
\bibitem{Weiss}U. Weiss, {\em Quantum Dissipative Systems}, World
Scientific (Singapore) (1993)
\bibitem{Strunz}W. T. Strunz, {\em Coherent Evolution in Noisy Environments} pp.377-392,
A. Buchleitner and K. Hornberger (Eds.), Springer (2002)
\bibitem{Haug}H. Haug and A.-P. Jauho,
{\em Quantum Kinetics in Transport and Optics of Semiconductors},
Springer (1996)
\bibitem{Haake} F. Haake, {\em Quantum Signatures of
Chaos}, Springer (Berlin, Heidelberg) (2001)
\bibitem{Milburn}G. J. Milburn, Phys. Rev. A {\bf 44}, 5401(1991); J. Finkelstein
Phys. Rev. A 47, 2412 (1993); G. J. Milburn Phys. Rev. A 47, 2415
(1993)
\bibitem{My1}An Min Wang, ``{\em Quantum mechanics in general
quantum system (I)}: exact solution", preprint, quant-ph/0611216.
Its earlier version is quant-ph/0602055
\bibitem{My2}An Min Wang, ``{\em Quantum mechanics in general
quantum system (II): perturbation theory}", preprint,
quant-ph/0611217. Its earlier version is quant-ph/0602055
\bibitem{Kimm}K. Kimm and H. Kwon, Phys. Rev. A {\bf 65},
022311(2002)
\bibitem{Zurek1982}W. H. Zurek, Phys. Rev. D {\bf 26} 1862(1982)
\bibitem{Cucchietti}F. M. Cucchietti, J. P. Paz, and W. H. Zurek,
Phys. Rev. A {\bf 72}, 052113(2005)
\bibitem{Ourmd}X. S. Ma, A. M. Wang, X. D. Yang, and F. Xu, Eur.
Phys. J. D {\bf 37}, 135(2006)
\bibitem{Ournew}In preparing
\end{references}
\end{document}